\documentclass[manuscript,letterpaper]{emulateapj}
\usepackage{natbib}
\usepackage{graphicx}
\usepackage{float}
\usepackage[]{hyperref}

\begin{document}
\title{Flare ribbon energetics in the early phase of an SDO flare }

\author{L.~Fletcher,  I.~G.~Hannah, H~S.~Hudson \altaffilmark{1}}
\affil{School of Physics and Astronomy, SUPA, University of Glasgow, Glasgow, G12 8QQ, U. K.}
\and
\author{D.E. ~Innes}
\affil{Max Planck Institute for Solar System Research, Max-Planck-Str. 2, 37191 Katlenburg-Lindau, Germany}

\altaffiltext{1}{Space Sciences Laboratory, U. C. Berkeley, 7 Gauss Way, Berkeley, CA 94720, USA}

\begin{abstract}
   {The sites of chromospheric excitation during solar flares are marked by extended extreme ultraviolet ribbons and hard X-ray footpoints. The standard interpretation is that these are the result of heating and bremsstrahlung emission from non-thermal electrons precipitating from the corona. We examine this picture using multi-wavelength observations of the early phase of an M-class flare SOL2010-08-07T18:24.}
   {We aim to determine the properties of the heated plasma in the flare ribbons, and to understand the partition of the power input into radiative and conductive losses.}
   {Using GOES, SDO/EVE, SDO/AIA and RHESSI we measure the temperature, emission measure and differential emission measure of the flare ribbons, and deduce approximate density values. The non-thermal emission measure, and the collisional thick target energy input to the ribbons are obtained from RHESSI using standard methods.}
   {We deduce the existence of a substantial amount of plasma at 10~MK in the flare ribbons, during the pre-impulsive and early-impulsive phase of the flare. The average column emission measure of this hot component is a few times $10^{28}{\rm cm}^{-5}$, and we can calculate that its predicted conductive losses dominate its measured radiative losses. If the power input to the hot ribbon plasma is due to collisional energy deposition by an electron beam from the corona then a low-energy cutoff of $\sim$ 5~keV is necessary to balance the conductive losses, implying a very large electron energy content.}
   {Independent of the standard collisional thick-target electron beam interpretation, the observed non-thermal X-rays can be provided if one electron in $10^3 - 10^4$ in the 10~MK (1~keV) ribbon plasma has an energy above 10~keV. We speculate that this could arise if a non-thermal tail is generated in the ribbon plasma which is being heated by other means, for example by waves or turbulence.}
\end{abstract}

\keywords{Sun: activity - Sun: chromosphere - Sun: flares - Sun: UV radiation - Sun:X-rays, gamma rays}


\section{Introduction}

How does the solar chromosphere in a flare evolve from its quiescent, pre-flare state to being the source of intense radiation and the site of energy deposition by non-thermal particles during the flare impulsive phase? Many multi-wavelength observations of the impulsive phase illuminate both thermal and non-thermal properties of the lower atmosphere, but detailed observations of the early stages of a solar flare - clearly a crucial part of the overall event - are sparse.  In this paper, we aim to characterize the flare's initial heating and brightenings, and the ramp-up of particle acceleration. We present multi-instrument observations of SOL2010-08-07T18:24, an M-class event in which the early evolution of flare ribbons and loops can be followed in detail, and the temperature, emission measure (EM), electron and plasma energy content and luminosity of the flare ribbons can be inferred.

Pre-flare studies are rather rare, in part due to flare triggers on previous missions which tended to start high-cadence observations only once the flare region became comparatively bright. Work has also tended to concentrate on the temporal and spatial relationships between pre-flare and flare sources, rather than on the physical conditions in the early phase. For example, \citet{1996SoPh..165..169F} and \citet{1998SoPh..183..339F} could find no clear spatial relationship between pre-flare and flare sources observed with the {\emph{Yohkoh}} soft X-ray telescope, and using the Transition Region and Coronal Explorer and {\it Yohkoh}/HXT, \citet{2001ApJ...560L..87W} found that the first hard X-ray bursts of a flare tend to occur in ultraviolet  footpoints that show no activity before the flare hard X-rays start.  A recent study of the early emission from a GOES B4.8 flare with the the Atmospheric Imaging Assembly (AIA) on the Solar Dynamics Observatory (SDO) and
the Coronal Diagnostic Spectrometer (CDS) on the Solar and Heliospheric Observatory (SOHO) showed that in microflares
chromospheric emission precedes hot plasma emission by about 6 minutes. \citep{2012A&A...540A..24B}.
The spatial correspondence of pre-flare source locations and other activity signatures has also been investigated - for example \citet{2006A&A...458..965C,2007A&A...472..967C} examined soft X-ray (SXR), hard X-ray (HXR) and extreme ultraviolet (EUV) pre-flare brightenings in events exhibiting filament ejection, finding evidence for heating and electron acceleration up to 50 minutes before a flare which was interpreted as evidence for pre-flare reconnection leading to filament destabilisation and eruption.

In this paper, we use a combination of data from the Ramaty High Energy Solar Spectroscopic Imager \citep[RHESSI, ][]{2002SoPh..210....3L} and Solar Dynamics Observatory Extreme UV Variability Experiment;  \citep[EVE][]{2010SoPh..tmp....3W} and Atmospheric Imaging Assembly; \citep[AIA][]{2011SoPh..tmp..172L}, supported by GOES X-ray measurements, to quantify the thermal and non-thermal energy content and luminosity in early flare ribbons. The consistent dataset afforded by these instruments allows us to investigate the physical properties of the flare plasma during the first 10-15 minutes of its development.

\section{Overview of the event}

\begin{figure*}
\centering
  \includegraphics[width=1.0\textwidth]{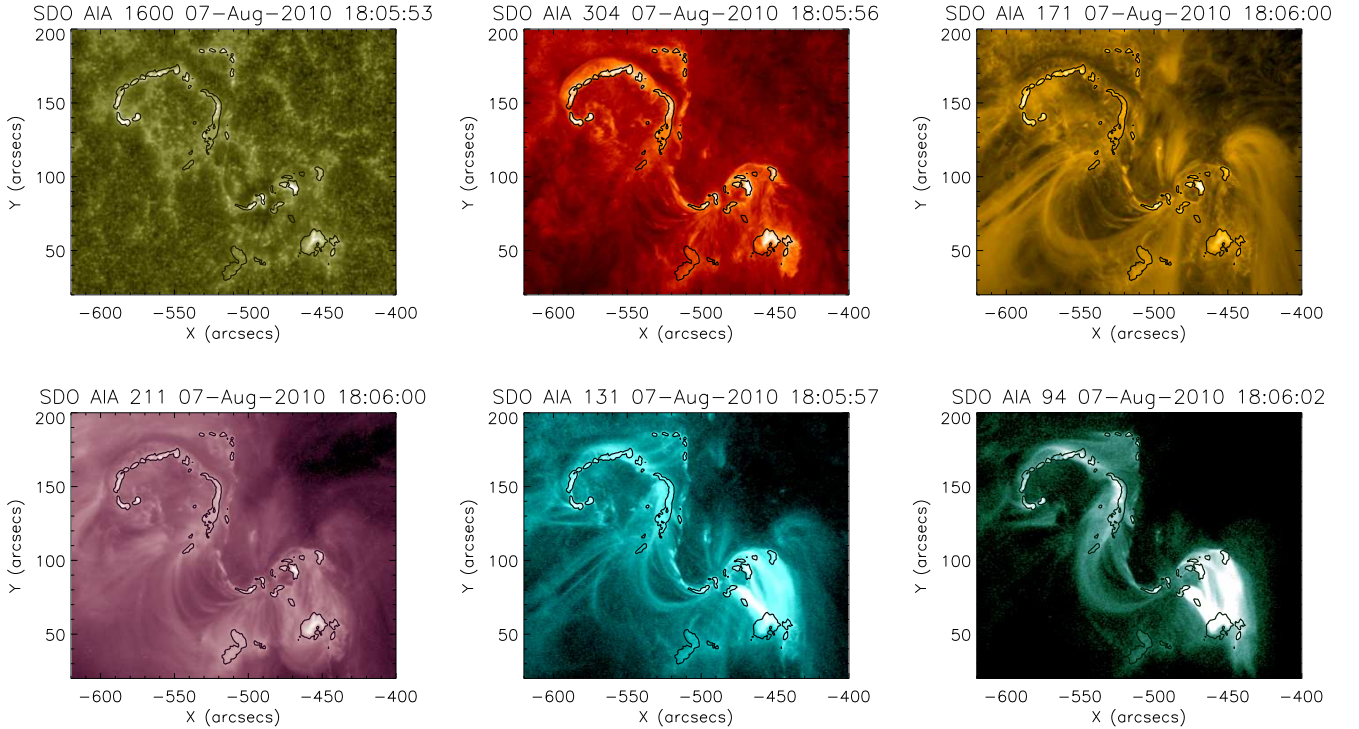}
     \caption{Almost simultaneous images of the flaring region in various SDO/AIA filters. 211~\AA\ channel contours at 700 ${\rm DN s^{-1}px^{-1}}$ are overplotted on all panels. Images have been logarithmically scaled.}
     \label{fig:aia6panelsample}
\end{figure*}

We will be concerned with the physical properties of the flare ribbons during the the first 10-15 minutes of the flare, rather than its evolution overall, but the event is interesting and merits a brief description. SOL2010-08-07T18:24 was an M1.0 class flare occurring in NOAA active region 11093 - the only major event reported from this region. The active region was the only one present on the solar disk at that time. It had one strong negative polarity sunspot in the south-east, and dispersed positive plage in the north-west. The plage hosted an inverse-S shaped filament, the southern part of which was apparently rooted in the sunspot. The flare was associated with the lifting-off of the upper portion of the filament, which began a slow rise at 17:40~UT, as studied in detail by \cite{2012SoPh..277..337V}. Prior to the flare, a set of loops rooted in the southern sunspot was very bright in EUV and X-rays (see Sections~\ref{sect:aiaimages} and \ref{sect:rhessi}), and dots of EUV emission were visible in the sunspot umbra and penumbra from 17:52~UT onwards. The flare ribbons started to become visible around 17:54~UT and appeared initially as elongated EUV brightenings along the northern part of the filament that remained along the polarity inversion line. These early ribbon brightenings were very close to, or perhaps even within, the dark filament material. The filament is clearly visible in the 304~\AA, 171~\AA\ and 211~\AA\ images in Figure~\ref{fig:aia6panelsample}, which shows logarithmically scaled SDO/AIA images, taken about half way through the rise phase. The northern flare ribbons are visible in all wavelengths, and the southern loops are present in the higher temperature filters, particularly the 94~\AA\ images.

As the upper portion of the filament accelerated away, the flare ribbons intensifed and spread rapidly outwards from the remaining filament material. We note in passing that the partial ejection of the filament is consistent with reconnection internal to the filament, or between its upper and lower parts \citep[e.g.][]{2009A&A...498..295T}. Dynamics in the decay phase of the event are reported in \cite{2012ApJ...744..173S}

\section{GOES data}
\begin{figure}
\centering
   \includegraphics[width=8cm]{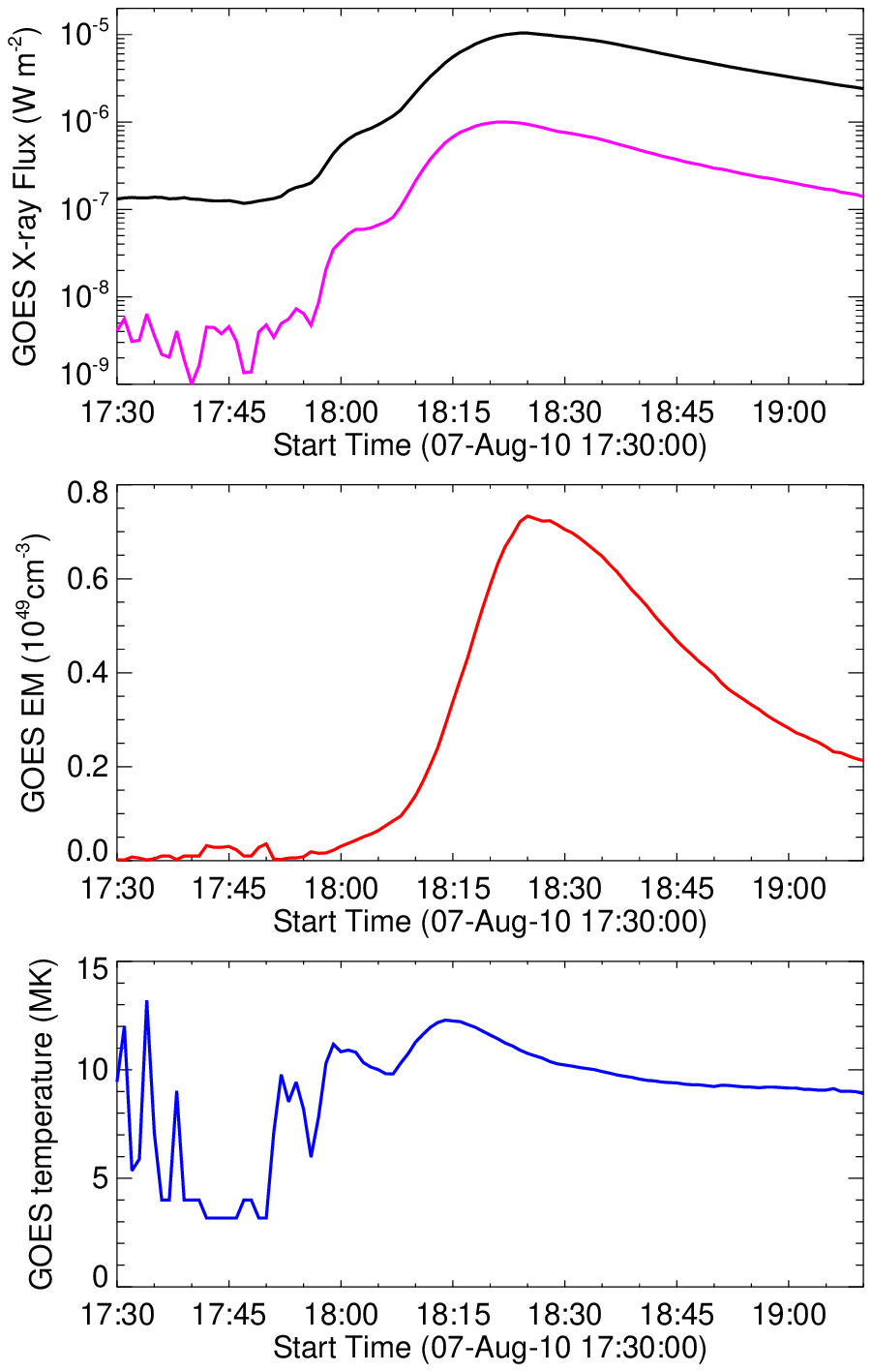}
     \caption{GOES X-ray flux (upper), EM (middle) and temperature (lower). In the upper panel the black line is the 1-8~\AA\ flux, and the pink line is the 0.4-4~\AA\ flux.}
     \label{fig:goes}
\end{figure}
The GOES lightcurve is shown in Figure~\ref{fig:goes}, along with the EM and temperature calculated with the GOES graphical user interface in the IDL SolarSoft library, having first subtracted the pre-flare background. This flare occurred in an otherwise completely quiet period so the background level is well-defined, and was taken between 16:00~UT and 17:00~UT. Data in the 0.5-4~\AA\ channel are noisy before around 17:55~UT, so the derived temperature and EM values are unreliable during this time. A rapid rise and bump in the GOES data between 17:56~UT and 18:05~UT is accompanied by a slow rise in the EM to a value of $6\times 10^{47}{\rm cm^{-3}}$, and an increase of the GOES temperature to 10~MK. Thereafter the EM rises rapidly to a peak of $7\times 10^{48}{\rm cm^{-3}}$, at 18:25~UT. As we will show in Section~\ref{sect:aialightcurves} the first increase in EM and temperature is associated with the southern loops, and after $\sim$ 18:06~UT the flare ribbons start to dominate.

\section{SDO data overview and analysis}

\subsection{SDO/EVE spectra and lightcurves}\label{sect:evegoes}

The MEGS-A spectrometer, one component of the EVE instrument \citep{2010SoPh..tmp....3W} covers the wavelength range 64 - 370~\AA, at a resolution of about 1~\AA, with 0.2~\AA\ binning in wavelength and 10~s binning in time \citep{2007SPIE.6689E..18C,2010SoPh..tmp...37H}. The instrument sees the whole Sun, and the flare emission is a small perturbation on a strong background. However, the spectra are very stable so the pre-flare spectrum can be subtracted to leave the flare enhancement. We will use the SDO/EVE to estimate contributions to the total flare luminosity, and as a quick way to examine the time evolution of the flare in each of the AIA passbands used for imaging.

\begin{figure*}
\centering
\hbox{
 \includegraphics[width=9.15cm]{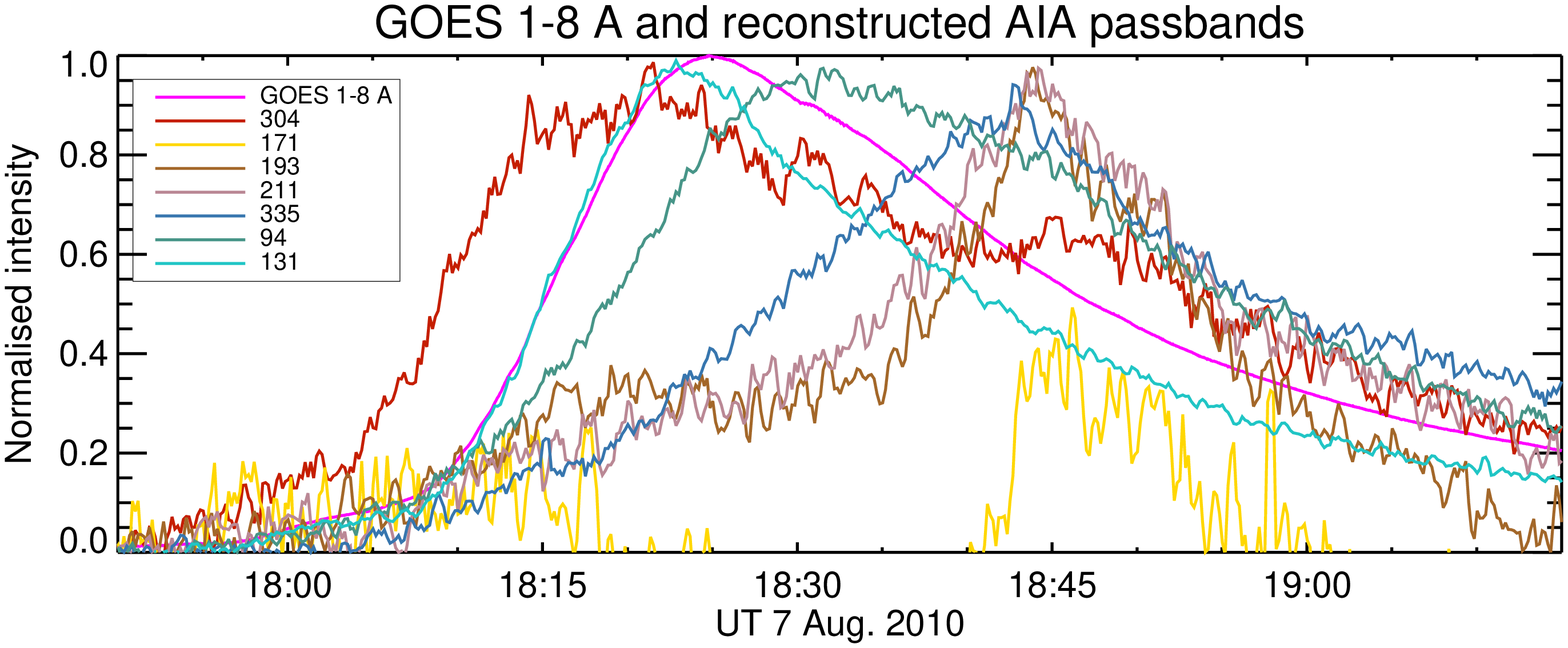} 
 \includegraphics[width=5.1cm]{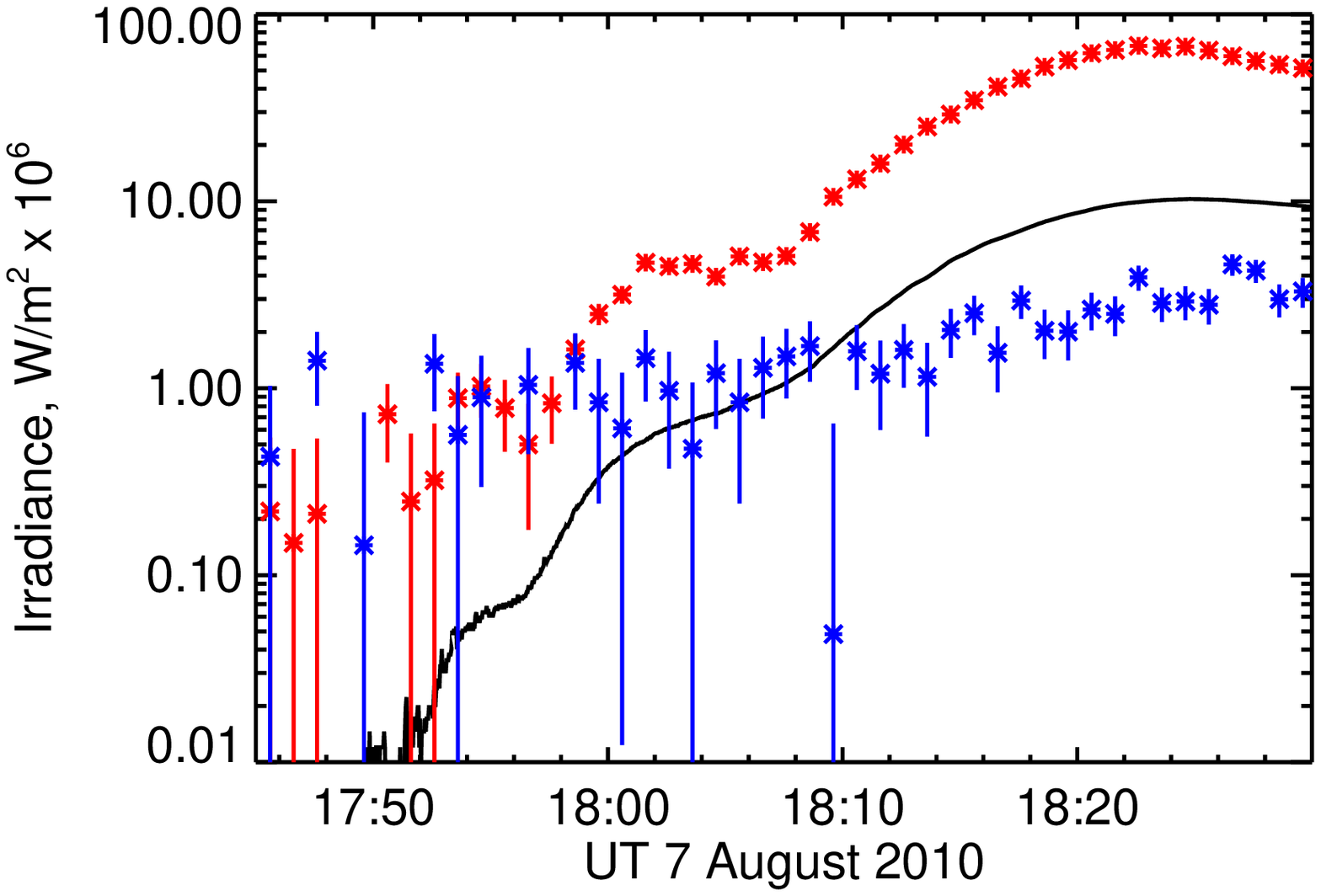} 
  }
     \caption{Left: normalized lightcurves of GOES 1-8~\AA\ and synthesized AIA intensities from EVE spectra. The synthesized AIA intensities have been smoothed with a 3-point boxcar.  Right: the GOES 1-8~\AA\ lightcurve and the flare excesses in the two primary lines featuring in the AIA/131\AA\ channel, i.e. Fe~{\sc{XXI}} at 128.75~\AA\ (red) and the Fe~{\sc{VIII}} pair at 130.94~\AA\ and 131.24~\AA\ (blue,  intensity upper limits).
     }
     \label{fig:EVE_synth}
\end{figure*}

The response in SDO/AIA can be simulated by convolving the EVE spectrum with the SDO spectral response \citep{2012SoPh..275...41B}. This is possible for all of the AIA EUV filters, and also the 304~\AA\ filter. The SDO/AIA EUV filters are narrow and centred on certain strong emission lines of iron, but these lines are superposed on a time-varying continuum \citep{2012ApJ...748L..14M}, which can also be strong, particularly in very hot events \citep[see e.g.][]{1999ApJ...511L..61F,2001ApJ...554L.103W}. Figure~\ref{fig:EVE_synth} (left panel) shows lightcurves obtained by convolving the observed EVE spectrum with the AIA filter responses. The synthesized 304~\AA\ lightcurve rises earliest of all the emissions plotted, as would be expected from modest excitation of the lower atmosphere. This corresponds to brightening of the chromospheric ribbons.

During the flare rise phase, the (normalized) synthesized 131 \AA\ channel lightcurve tracks the (normalized) GOES 1-8~\AA\ flux best of all the AIA passbands, suggesting a significant high temperature plasma contribution in the 131~\AA\ channel. 
 \citet{2010A&A...521A..21O} calculated contributions to the AIA EUV passbands using a flare prescribed DEM, albeit one representing the decay phase of an M2 flare \citep{1979ApJ...229..772D}, and showed that the 131~\AA\ channel is expected to be dominated by Fe~{\sc{XXI}} at 128.75~\AA\, formed at a temperature of 11~MK. However, there is also a low temperature contribution in this passband, dominated by the Fe~{\sc{VIII}} pair at 130.94~\AA\ and 131.24~\AA\ formed at 400,000~K \citep[See also][]{2012A&A...540A..24B}. These are not expected to be strong, but we check this by using EVE to measure the flaring excess integrated over the Fe~{\sc{VIII}} and Fe~{\sc{XXI}} lines. This is plotted in the right hand panel of Figure \ref{fig:EVE_synth}, which compares the absolute irradiance in GOES 1-8~\AA\ and the excess irradiance in  the Fe{~\sc{XXI}} line (red) and the Fe{~\sc{VIII}} pair (blue).  Note, at the wavelength position of the Fe{~\sc{VIII}} pair there is no spectral feature apparent; the counts are consistent with the background level. So the blue points plotted are mostly indicating the background continuum intensity, and should be considered only as upper limits to the Fe{~\sc{VIII}}  line intensity.  The excess is calculated with respect to a 5-minute quiet-Sun average irradiance taken before the flare starts. It is clear from its intensity and its evolution that the hot  Fe~{\sc{XXI}}-emitting plasma is the main contributor to the AIA/131~\AA\ channel throughout the flare. This is consistent with the temperatures derived from the GOES channel ratio in Figure~\ref{fig:goes}. 

We note also that the intensities in the other spectral passbands peak well after the GOES peak, consistent with a plasma cooling from high to low temperatures. The moderate-temperature channels (171~\AA, 193~\AA\ and 211~\AA) show intensity decreases after 18:15~UT, which may be consistent with a low-temperature coronal dimming. However, as the focus of this paper is on the flare rise phase we will not consider the late-phase development any further.


\subsection{SDO / AIA imaging}\label{sect:aiaimages}
We use AIA level 1 data, converted to level 1.5 data using the SSW \verb1aia_prep1 routine, and normalized by the exposure time. The data have been despiked, and inspection of the despiked pixel locations shows that only around 10 flaring pixels per image have been removed over the time of interest. The main flare ribbons in the north start to appear around 17:54~UT, and are well-defined, mostly unsaturated, and easy to distinguish from the main loop arcade that starts to brighten from around 18:10~UT onwards. The location of ribbons is compared across AIA channels in Figure \ref{fig:aia6panelsample}. Contours at 
700 ${\rm DN s^{-1}px^{-1}}$
 from the 211~\AA\ filter are overlaid on the other images. There is a very good correspondence - at the level of one or two pixels -  between the brightest ribbon sources in the north across the AIA wavelength range. Several other sources in the southern sunspot area are also spatially well-correlated. This indicates that the ribbons emit across a broad range of temperatures, from around 100,000~K to $\sim 10$~MK.

\subsection{AIA 131~\AA\ and GOES lightcurves}\label{sect:aialightcurves}

 \begin{figure*}
 \centering
  \includegraphics[width=\textwidth]{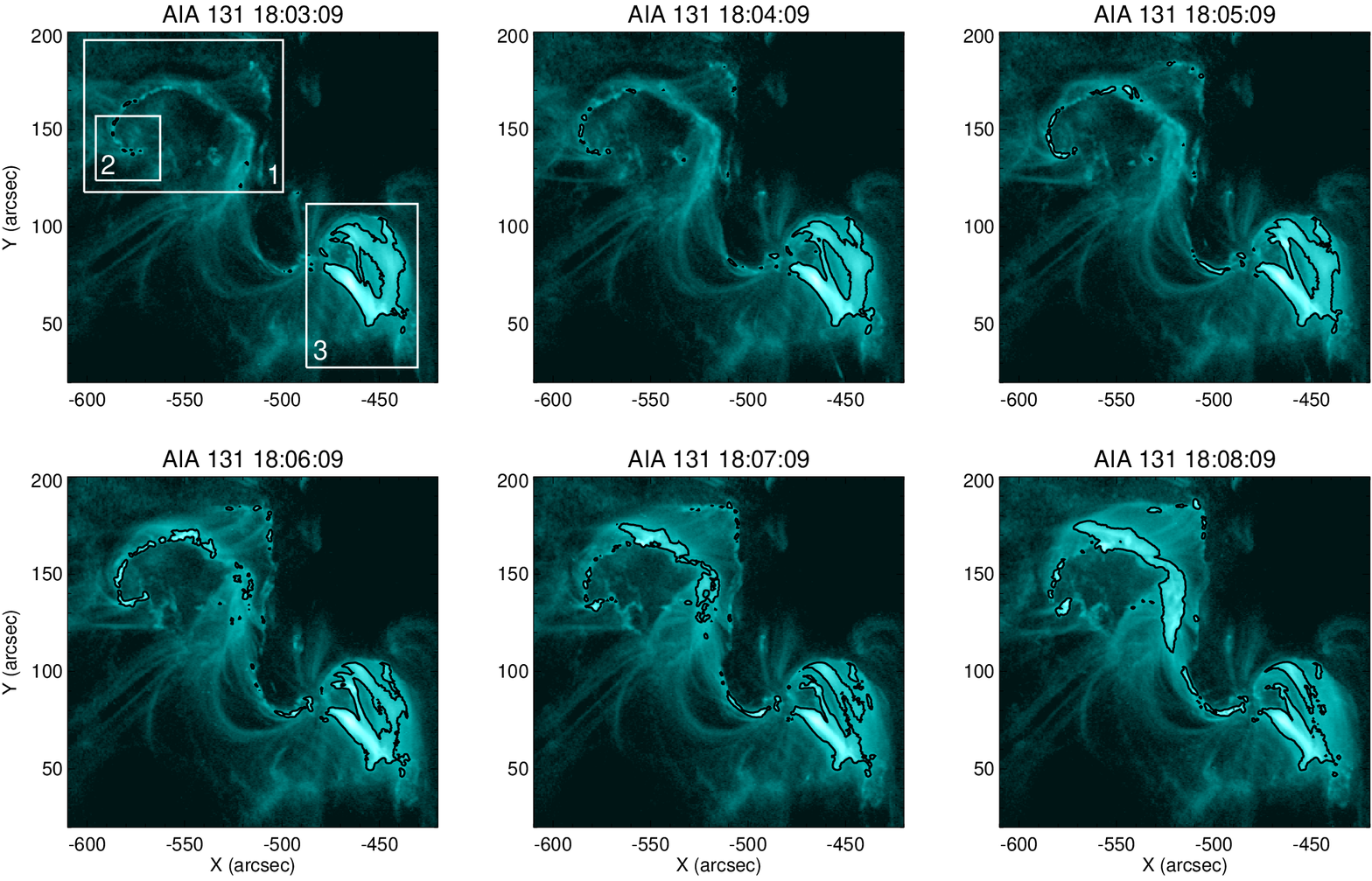}
 \caption{SDO/AIA 131\AA\ channel images, with contours drawn at 175 ${\rm DN s^{-1}px^{-1}}$. Summing inside these contours results in the lightcurves in Figure~\ref{fig:AIA131curves}. Boxes in the top left image outline three different regions of interest: (1) northern ribbons; (2) subset of these ribbons, referred to as the northern footpoints; (3) southern loops. The image intensity has been logarithmically scaled.}
\label{fig:131contours}
 \end{figure*}

The two main parts of the flare - the northern ribbons and the southern loops - have quite different characteristics, and the overall lightcurve in the AIA passbands is a composite of these. In Section \ref{sect:evegoes} we demonstrated that the synthesized 131~\AA\ lightcurve most closely matches the GOES lightcurves, and with SDO/AIA imaging the 131~\AA\ lightcurve can be decomposed into contributions from the different regions shown in Figure ~\ref{fig:131contours}. We separate the flare into three regions: the bright loops near the sunspot, at around (-470", 70"), the ribbons and loops to the north-east of these, and then a subset of pixels in the ribbons (referred to as the `northern footpoints') which are not so affected by the presence of loops later in the event. These regions are shown in Figure~\ref{fig:131contours}. We sum emission from sources within each of these three boxes having intensity above 175 ${\rm DN s^{-1}px^{-1}}$, as shown by the contours in Figure ~\ref{fig:131contours}. These capture the majority of the ribbons and loop emission.

\begin{figure}
\centering
\vbox{
  \includegraphics[width=8.5cm]{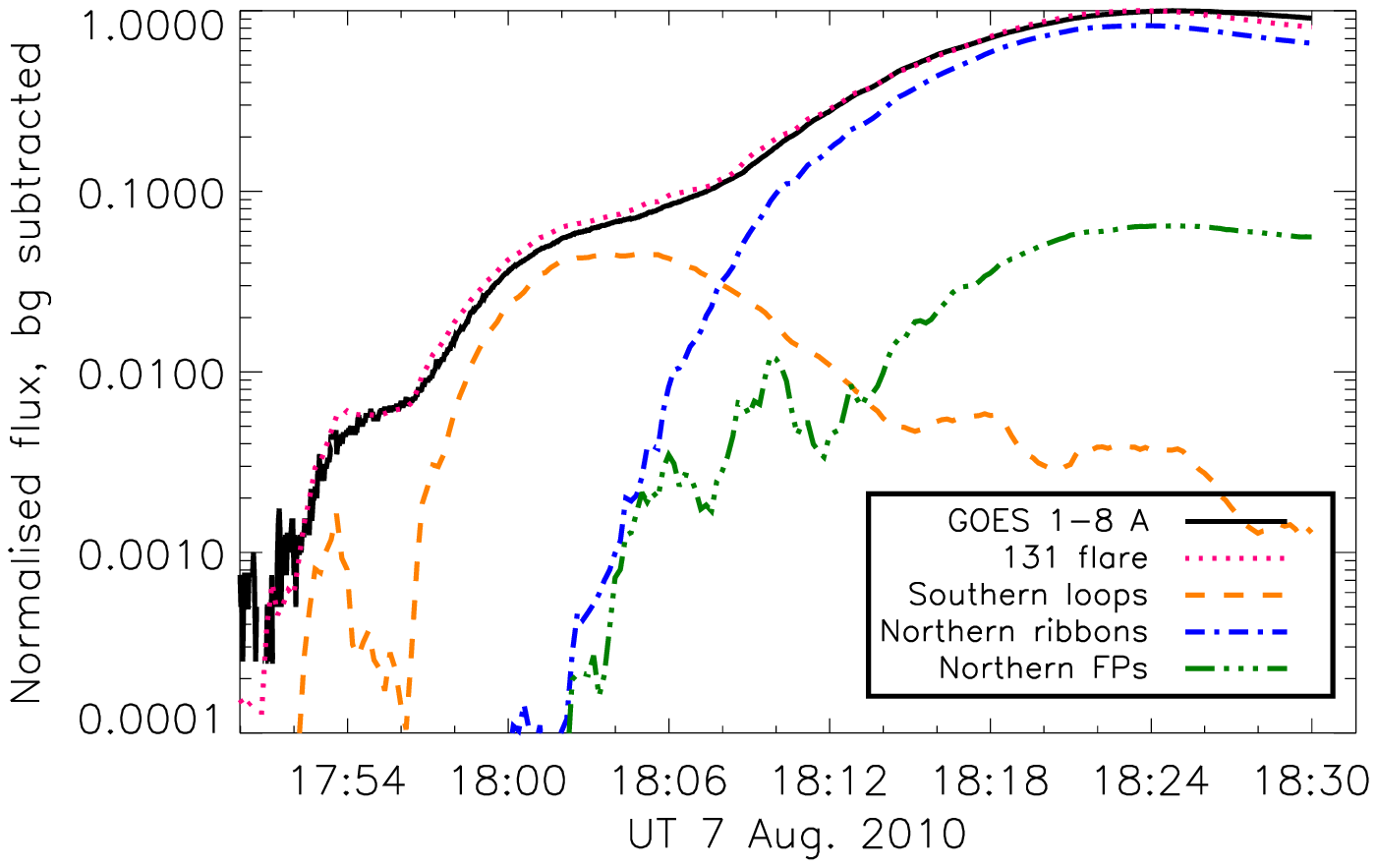}  
  \includegraphics[width=8.5cm]{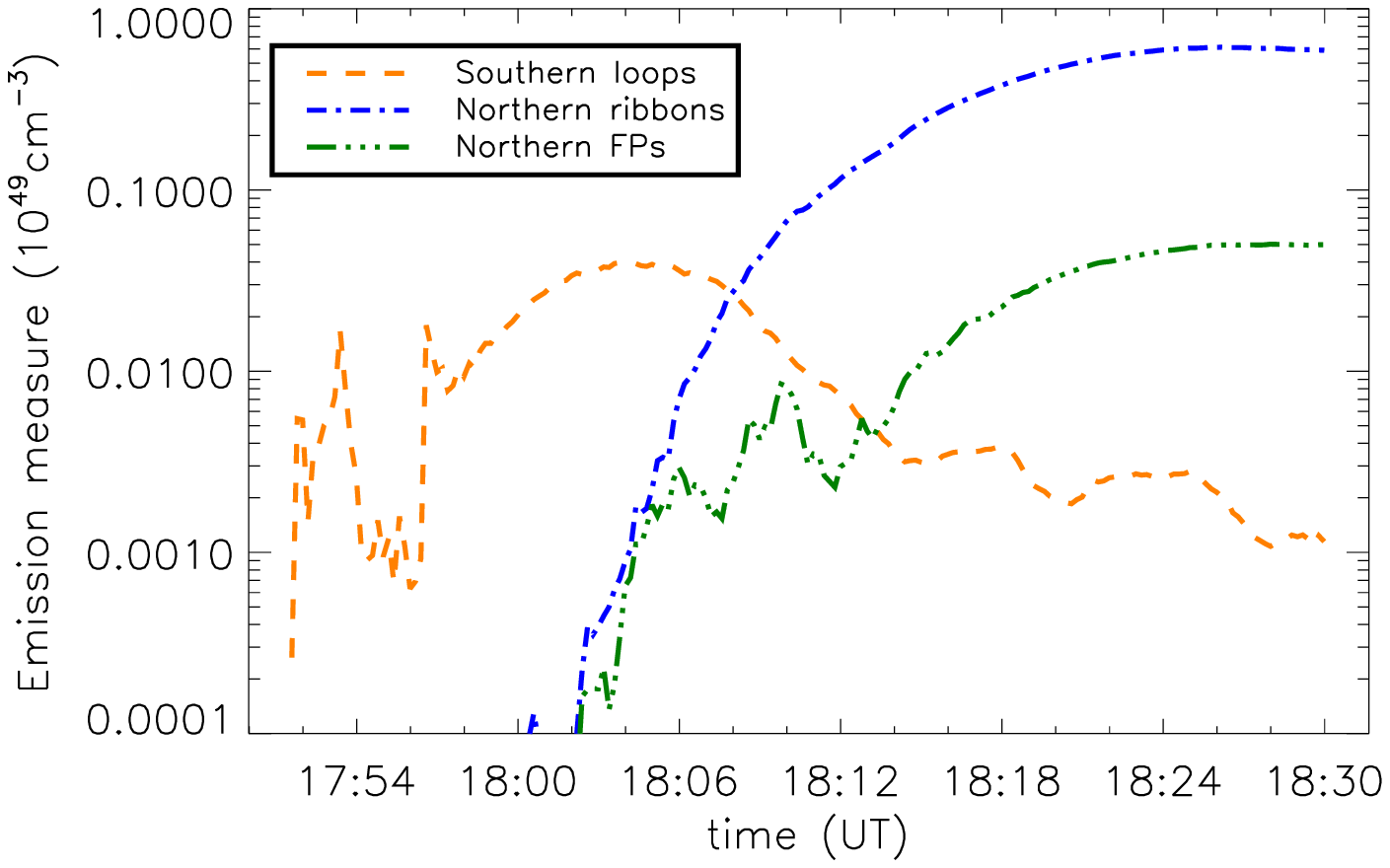}
  \includegraphics[width=8.5cm]{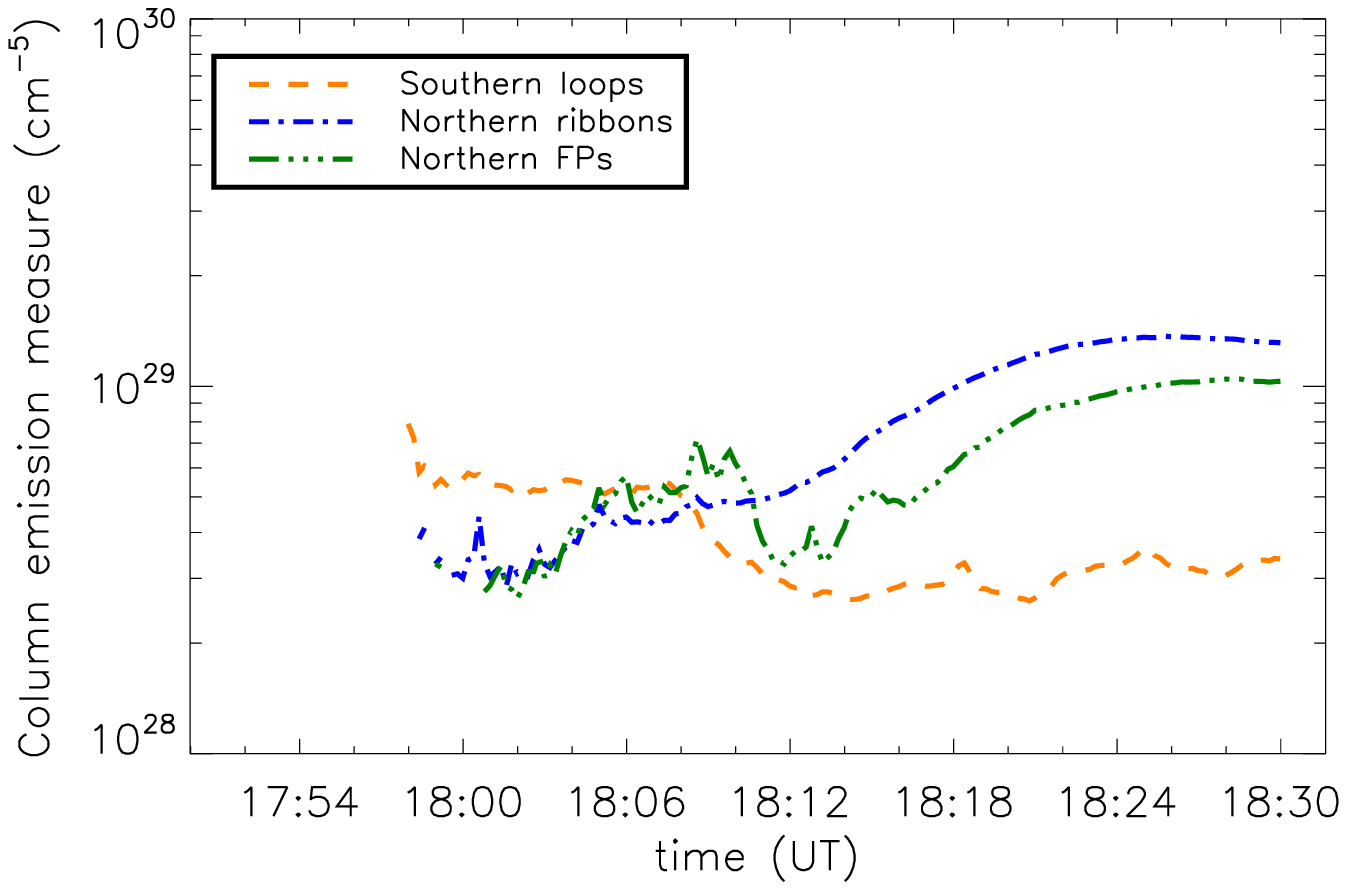}
 }
     \caption{Upper panel: Background-subtracted and normalized (to peak values) GOES 1-8~\AA\ flux (black solid curve) and SDO/AIA 131~\AA\ intensity of sources above 175 ${\rm DN s^{-1}px^{-1}}$ summed over the whole field-of-view of Figure~\ref{fig:131contours} (pink dotted line), the southern loops ( orange dashed line), northern ribbons (blue dot-dashed) and the subset of northern footpoints (green dash-triple-dot line).  Center panel: volume EMs inferred from the 175 ${\rm DN s^{-1}px^{-1}}$ thresholded sources in each of the boxes outlined in Figure \ref{fig:131contours}. Lower panel: column EMs. After 18:10~UT the integrated emission is dominated by loops}
     \label{fig:AIA131curves}
\end{figure}

The summed pre-flare background intensity from each box is subtracted from the summed intensity in sources above 175 ${\rm DN s^{-1}px^{-1}}$, and the result is normalized to the maximum value of the total background-subtracted 131~\AA\ intensity from the whole field-of-view. This results in the dotted, dashed and dot-dashed curves in the upper panel of Figure ~\ref{fig:AIA131curves}. The normalized and background-subtracted GOES 1-8~\AA\ emission is shown as a solid curve. Between around 17:58~UT and 18:08~UT the southern loops dominate the 131~\AA\ intensity. The emission from the northern regions, which include ribbons and later on loops, becomes significant ($\sim$ 10\% level) at 18:05~UT and starts to dominate the southern sources after 18:08~UT.  We conclude that there is a period from about 18:08 to 18:10 during which the AIA 131~\AA\ lightcurves, and by implication the GOES lightcurves, have a strong or even dominant contribution from the flare ribbons and footpoints. After around 18:10 the emission labeled  `northern ribbons'  in~\ref{fig:AIA131curves} develops a strong contribution from the loops which start to appear in the corresponding box. In the `northern footpoints' box, however, loops do not appear until later and the bursty character of the emission that one would associate with footpoints or ribbons rather than loops is apparent for longer - up to around 18:12. We reason that this bursty behaviour would be visible in the northern ribbons more generally, were it not masked by the appearance of the loops.

Since the normalized AIA 131~\AA\ emission tracks the GOES 1-8~\AA\ emission very closely, we may reasonably assume that on the whole they originate from the same plasma. It should be noted, however, that the 131~\AA\ band contains strong Fe{\sc{VIII}} lines as well as the Fe{\sc{XXI}} lines, so in Section~\ref{sect:dem} we employ a differential emission measure (DEM) analysis to further investigate the plasma temperature.
Multiplying the GOES EM by the fractional 131~\AA\ intensity associated with the different resolved AIA sources provides an estimate of the EM associated with each, shown in the centre panel of Figure~\ref{fig:AIA131curves}. So for example between 18:06:00~UT and 18:12:00~UT, the small group of northern footpoints have an EM of between $\sim 3 \times 10^{46}$ and $8\times 10^{46} \rm cm^{-3}$, while the northern ribbons as a whole at 18:10~UT have an EM of $7\times 10^{47}{\rm cm}^{-3}$. Dividing through by the total area (at the Sun) of the pixels in each source box above the limit of 175 ${\rm DN s^{-1}px^{-1}}$ we arrive at an estimate of the column EM for the hot ribbon plasma, shown in the lower panel of Figure~\ref{fig:AIA131curves}. This varies increases from 3 to 7 $\times 10^{28}{\rm cm^{-5}}$, between 18:00 and 18:12~UT.

\subsection{Overlays of flare sources on SDO/HMI}
\begin{figure}
\centering
\vbox{
\includegraphics[width=0.45\textwidth]{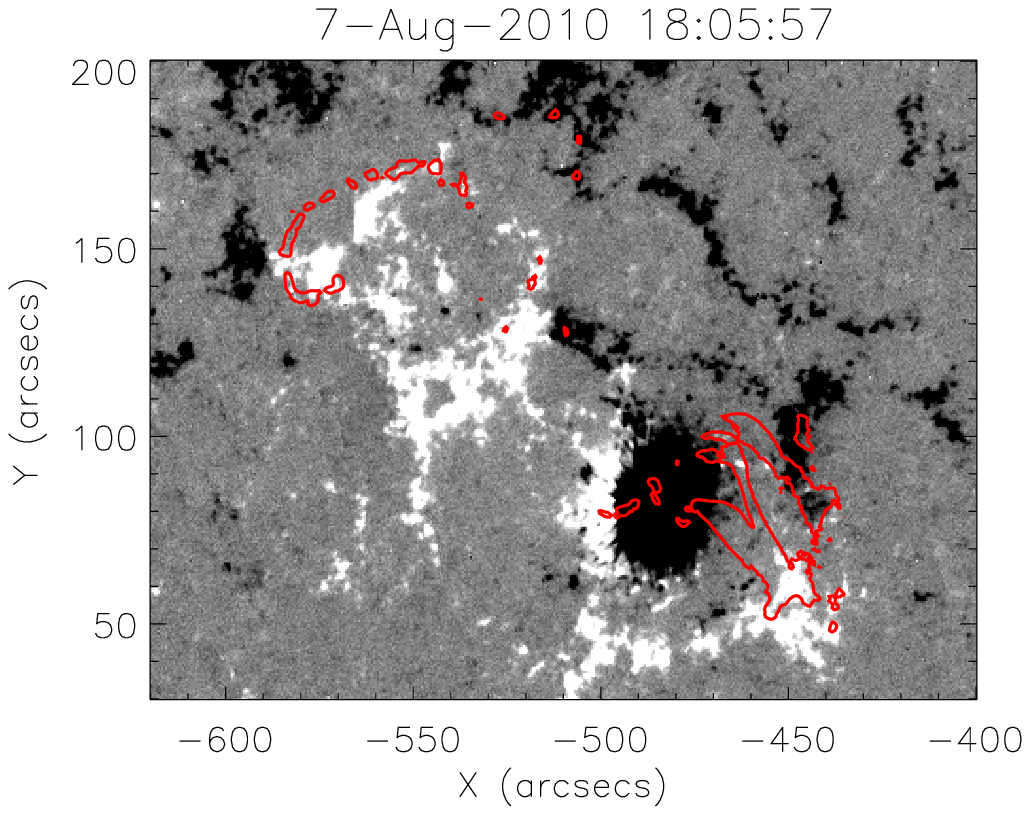}
\includegraphics[width=0.45\textwidth]{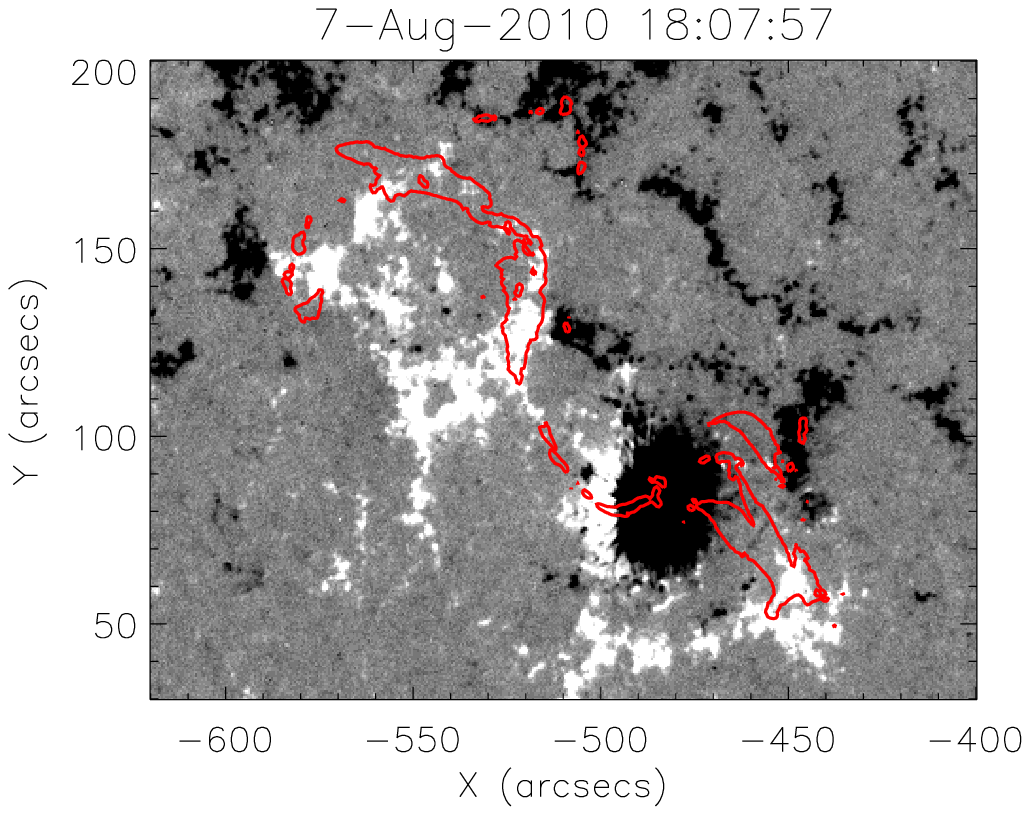}
}
\caption{Overlays of the 131~\AA\ flare sources on HMI line-of-sight magnetograms. Contours are at 200~DN/s.}
\label{fig:aia_hmi_overlays}
\end{figure}

Overlays of the 131~\AA\ flare sources on the nearest-in-time HMI line-of-sight magnetogram shown in Figure~\ref{fig:aia_hmi_overlays} demonstrate that the ribbons are mostly located in magnetic plage regions, but also that there are ribbon sources in weak-field regions, e.g. around (-570", 160"). These sources tend to be weaker in the EUV than their neighbours, and overlays with RHESSI (Fig~\ref{fig:RHESSI_ims}) also suggest an absence of RHESSI sources at this location.

\subsection{SDO / AIA differential emission measure}\label{sect:dem}

\citet{2012A&A...539A.146H} have developed a method of deriving the DEM from AIA observations, using a regularized inversion method. This permits an estimate to be made of the amount of plasma as a function of temperature at different locations in the flare, and provides estimates of both the DEM uncertainty and the effective temperature resolution. This method is applied to the three boxed regions highlighted in Figure~\ref{fig:131contours}, again selecting only pixels within the box which are above the 175 ${\rm DN s^{-1}px^{-1}}$ threshold. We have also concentrated on the time after 18:05~UT, when the ribbon emission is starting to become important. The source areas estimated from the 131~\AA\ images are used to convert the volume DEM into a column DEM.

\begin{figure*}
\centering
\includegraphics[width=0.9\textwidth]{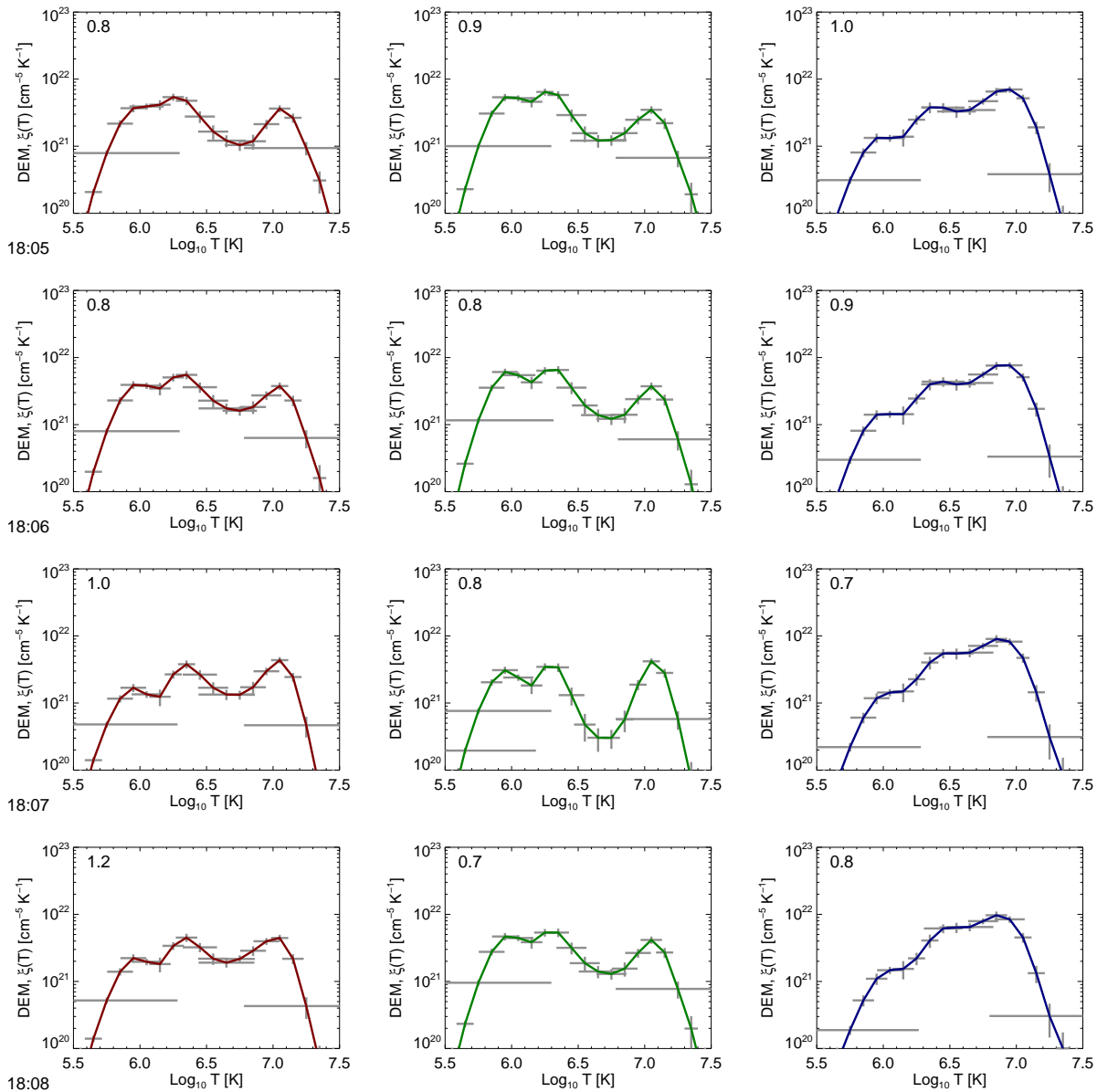}
     \caption{DEM from three different regions in the flare: left column - northern ribbons and (later) loops; centre column - northern footpoints; right column - southern loops. Each row corresponds to the time indicated underneath. These have been calculated for the default Chianti pressure of $10^{15}\rm{cm^{-3} K}$. Note that the temperatures with very large horizontal error bars do not just indicate a bad temperature resolution but a poorly recovered DEM solution at those points \citep{2012A&A...539A.146H}. The $\chi^2$ between the data and the recovered DEM is indicated in the top-left corner of each panel.}
     \label{fig:AIA_DEM}
\end{figure*}

\begin{figure*}
\centering
\includegraphics[width=0.9\textwidth]{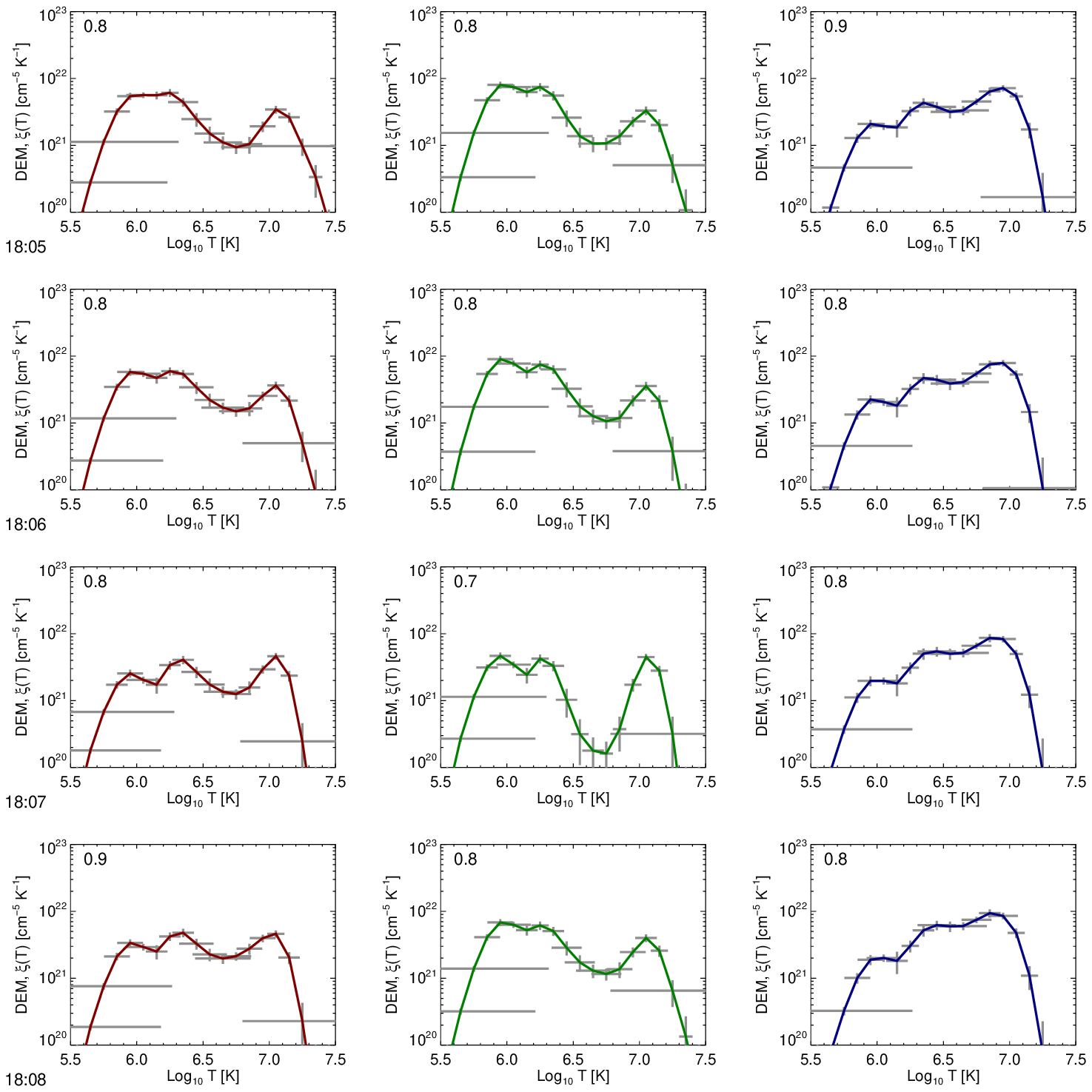}
     \caption{As Figure~\ref{fig:AIA_DEM} but calculated for a pressure of $10^{18}\rm{cm^{-3} K}$.}
     \label{fig:AIA_DEM2}
\end{figure*}

The DEMs are made using the 6 AIA EUV channels, and the accuracy of the results depends on our knowledge of the spectral lines in the channels. As shown by simultaneous spectroscopic and imaging observations with the 
Hinode EUV imaging Spectrometer (EIS) and SDO/AIA, several of the observed transition region lines are omitted or poorly represented by CHIANTI \citep{2011A&A...535A..46D}. In Figures~\ref{fig:AIA_DEM} and ~\ref{fig:AIA_DEM2} we show results using the AIA temperature response functions \citep{2012SoPh..275...41B} calculated at the default pressure in Chianti of $10^{15}\rm{cm^{-3} K}$, and also using contribution functions calculated using a greatly increased pressure of $10^{18}\rm{cm^{-3} K}$ which may be appropriate for flaring conditions. We also include the recent normalizations implemented to provide good agreement with SDO/EVE; empirical corrections to the 94~\AA\ and 131~\AA\ channels compensate for the omissions in the CHIANTI database \citep{2012ApJ...745..111T}.  With these corrections the response of SDO/AIA to high and low  temperature flare emission is believed to be accurate to 25\% \footnote{\url{http://sohowww.nascom.nasa.gov/solarsoft/sdo/aia/response/chiantifix_notes.txt}}. These are however empirical corrections and  improvements to the CHIANTI atomic database are required and currently  being studied.

 DEM analysis proceeds under assumptions of thermal and ionization equilibrium, and of optically thin radiation, and we must consider the validity of these assumptions under flaring conditions. A more detailed discussion is available in \cite{2013arXiv1302.2514G} but, briefly stated, the assumptions of thermal and ionization equilibrium coupled with a slow flare evolution in the early phase means that, even at these high flare temperatures, equilibrium assumptions are likely to be adequate in flare footpoints. This is because the footpoint density is relatively high; we estimate $n_e \sim 10^{10}{\rm cm}^{-3}$ (see Section~\ref{sect:balance} and  \cite{2011A&A...532A..27G}). High density ensures that the core of the electron distribution reaches thermal equilibrium quickly, even at high temperatures, unless there is a very short heating timescale. In the case of slow heating departure from ionization equilibrium is small for densities above $10^{10}{\rm cm}^{-3}$ \citep{2009A&A...502..409B}.   Low opacity in all of the main contributing lines formed at temperatures greater than  $10^6$~K  can be verified using oscillator strengths from the Chianti database and Equation (1) from \cite{2002A&A...390..219B}. The Fe~{\sc{IX}} line at 171.07~\AA\ that dominates the AIA/171 channel has a large oscillator strength, and thus a larger opacity. However even under the most pessimistic assumption (of coronal abundances) its total optical depth is less than unity for column depths under $\sim 10^{18}{\rm cm^{-2}}$, and smaller still if photospheric abundances are used. We conclude that the assumption of optically thin emission is reasonable in the footpoints of this flare, even given the rather dense plasmas we find.

The DEMs calculated for the two values of pressure are similar in shape and in magnitude, with variations everywhere less than a factor two. At  $p = 10^{18}\rm{cm^{-3} K}$, the DEM is slightly larger at log~T~$\sim 6$ and slightly smaller at log~T~$\sim 6.5 - 7$ than the corresponding value at $p = 10^{15}\rm{cm^{-3} K}$. Each of the DEMs shown in Figure~\ref{fig:AIA_DEM} has three significant bumps, at around log~T~$\sim 5.9-6$, log~T~$\sim 6.3-6.4$ and log~T~$\sim 7.0-7.1$. At the lower temperatures it is expected that there will also be a contribution from the ambient coronal plasma in the line-of-sight between the observer and the flare sources. The largest values of high-temperature DEM are found in the southern loops, but the ribbons and footpoints also have substantial amounts of high-temperature plasma. There is some evidence that, over the 4 minutes examined, the amount of plasma at  $10^6$ to $10^{6.3}$ ~K plasma in the ribbons decreases while the amount of $10^7$~K plasma stays relatively constant. However, the uncertainties on the DEM recovery make it hard to be definitive about this. There does appear to be a significant decrease in the EM at around $\log~T = 6.7$ in the northern footpoints at 18:07~UT. We note this, but do not currently have an explanation. It may be a numerical instability in the reconstruction, though the error bars look very similar to the other cases. It is unlikely to be due to the loss of material from the region with the coronal mass ejection, because the DEM is generated just from the footpoint sources above some threshold intensity. Any decrease in the amount of material at $\log~T = 6.7$ due to evaporation would presumably also lead to a decrease in the overlying $\log~T = 7$ material, which is not seen. A pixel-by-pixel examination of the EM distribution, which we plan for a future paper, may shed light on this.

The units of the DEM are $\rm{cm}^{-5}\rm{K}^{-1}$, and an estimate of the EM in units of $\rm{cm}^{-5}$ at each temperature can be obtained by multiplying by that temperature, meaning that the peak at $10^7$~K has the largest EM contribution, of $\sim 5 \times 10^{28}\rm{cm}^{-3}$. This is consistent with the EM values obtained from GOES at $10^7$~K. We note also that EIS spectroscopy has been used in other events to demonstrate the presence of substantial amounts of 10~MK footpoint plasma \cite{2013arXiv1302.2514G}.

\section{RHESSI spectroscopy and imaging}\label{sect:rhessi}
\begin{figure}
\centering
\includegraphics[width=0.45\textwidth]{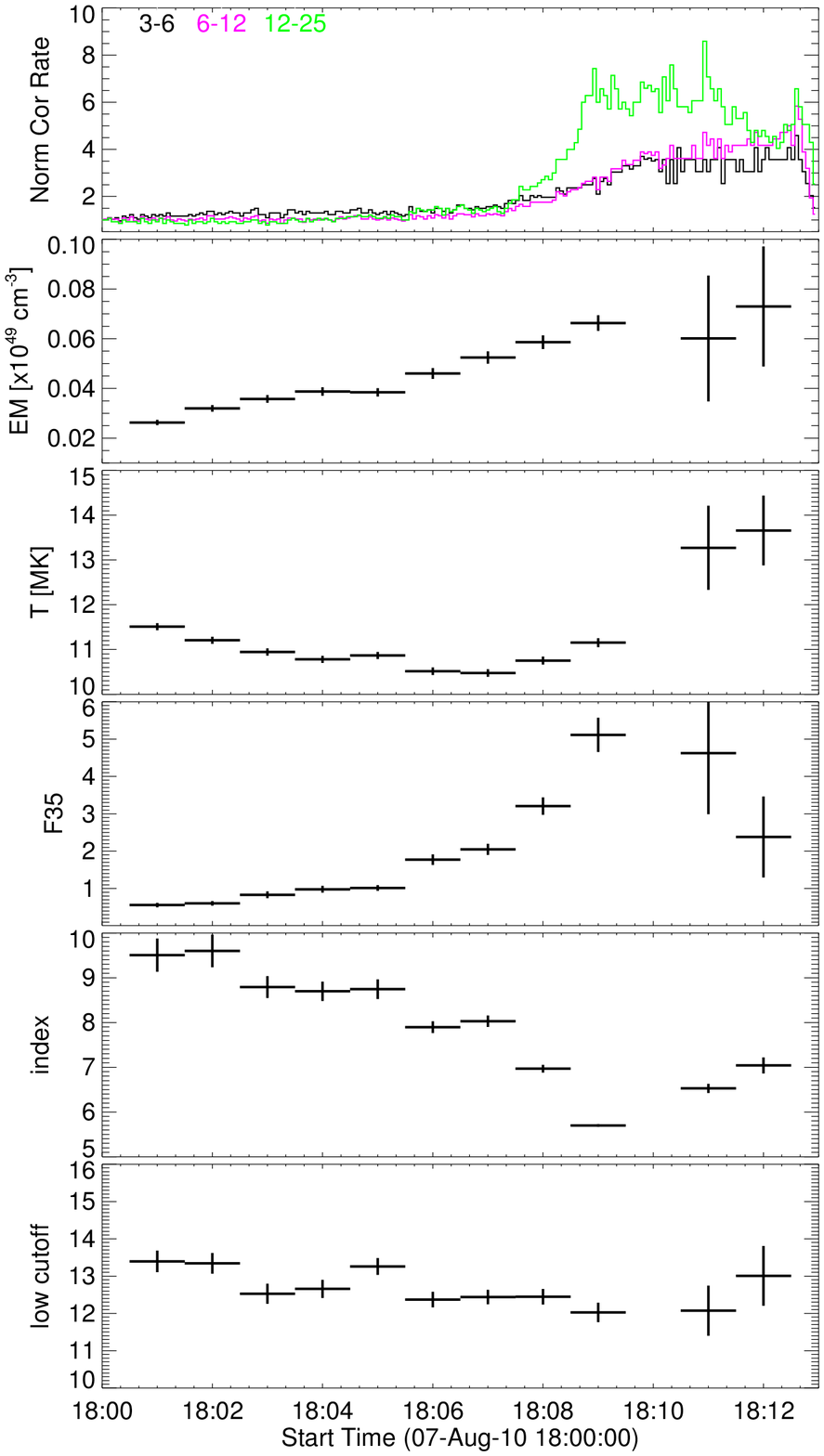}
\caption{Evolution with time of the RHESSI spectral fit parameters. From the top, the panels show the normalized corrected RHESSI count rates in 3-6, 6-12 and 12-25keV, the EM of the isothermal plasma component, the temperature of this component, the total non-thermal electron number flux, $F_{35}$ above low-energy cutoff $E_c$ in units of $10^{35}$ electrons~s$^{-1}$, the electron spectral index $\delta$ and the electron low-energy cutoff.}
\label{fig:rhessi_spec}
\end{figure}

The time between 18:00:30~UT and 18:12:30~UT is split into intervals of 60 seconds, and RHESSI spectra are fitted in electron space, with an isothermal plus single power-law fit, using the OSPEX spectral fitting software. The fit assumes that the non-thermal emission originates from a cold, collisional thick target. The thermal component is parameterized by a single temperature and EM. The non-thermal electron spectrum is parameterized by the total number of electrons per second above the low-energy cutoff, $E_c$, and the electron spectral index $\delta$. The fit results are shown in Figure~\ref{fig:rhessi_spec}. No fit parameters are shown between 18:09:30~UT and 18:10:30~UT, as this coincided with change in the attenuator state. The spacecraft entered night at 18:12~UT.

The RHESSI normalized countrates, shown in the upper panel of Figure~\ref{fig:rhessi_spec}, are very stable for the first few minutes of the flare. At 3-6 and 6-12~keV the countrate is almost constant between 18:00:30~UT and 18:06:30~UT, followed by a smooth increase in intensity. Images at this time show that the emission is dominated by the hot southern loops, with no ribbon emission visible at the 10\% level. The 12-25 keV intensity increases gradually from around 18:04~UT. Between 18:00:30~UT and 18:10:30~UT the RHESSI EM increases gradually from $2.5$ to $5\times 10^{47}\rm{cm^{-3}}$ (panel 2 in Figure~\ref{fig:rhessi_spec}). Since the AIA analysis in Section~\ref{sect:aialightcurves} indicates that from 18:06~UT the southern loop EM is decreasing, the increase in the RHESSI count rates and EM after this time, and up until 18:10~UT is most likely attributable to the northern ribbons. The RHESSI EM varies from $4-7\times 10^{47}{\rm cm}^{-3}$ between 18:06 and 18:09~UT, which is broadly consistent with the values obtained from GOES. The temperature (Fig~\ref{fig:rhessi_spec} panel 3) decreases slightly from 11.5~MK to 10.5~MK at 18.07~UT before rising again, possibly also due to the brightening of the northern ribbons.

The spectrum, which starts off very soft, hardens abruptly between 18:06~UT and 18:10~UT as the ribbons brighten (Fig~\ref{fig:rhessi_spec} panel 5). This is reflected also in the increasing electron flux during this time (panel 4) and the decrease in spectral index from 8 to 6. The fitted low-energy cutoff remains approximately constant throughout, at around 12-13~keV (panel 6).

\begin{figure*}
\centering
\includegraphics[width=\textwidth]{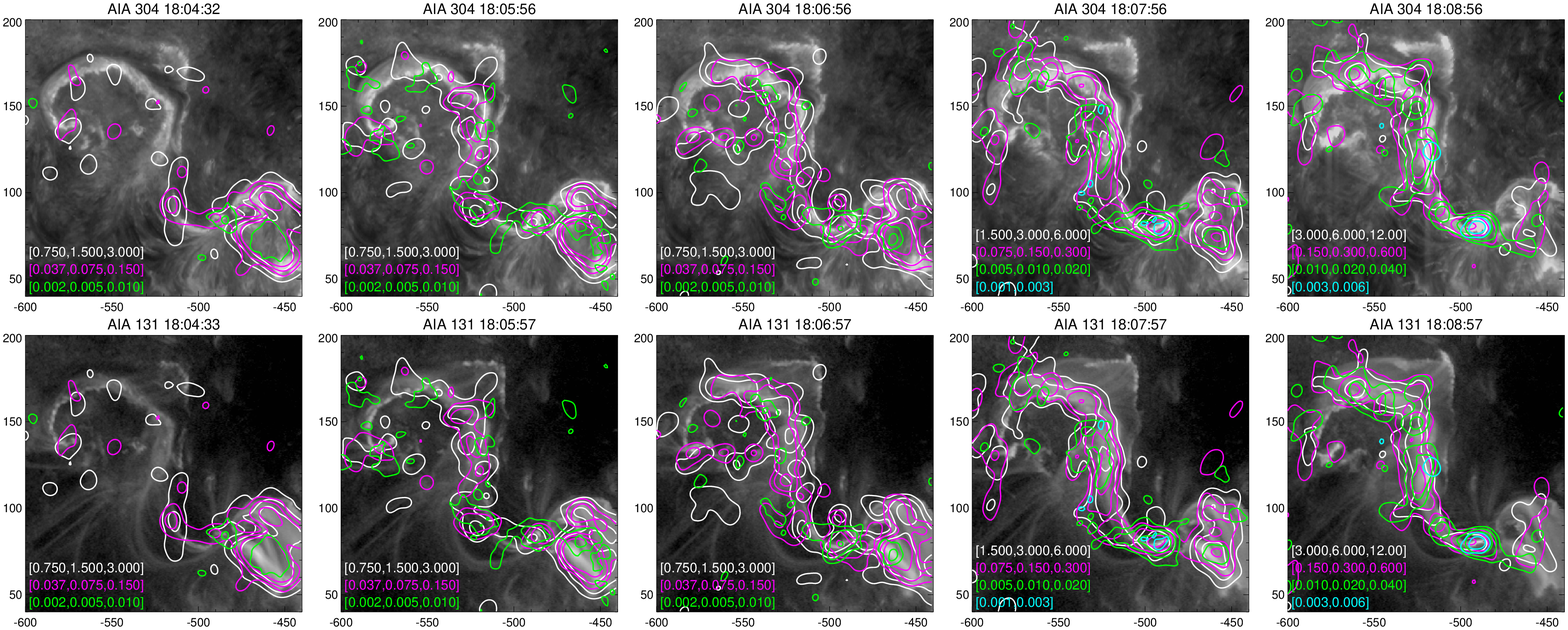}
\caption{RHESSI images made using the Clean algorithm and data from detectors 2-8. Contours plotted are absolute values, in counts/s/detector/keV as shown in the lower left of each panel. In the first 3 columns, contours shown all have the same absolute values between images. Contours are 3-6~keV (white); 6-12~keV (pink); 12-25~keV (green); and in the rightmost 2 columns also 25-50~keV (cyan). Axes are arcseconds from disc centre.}
 \label{fig:RHESSI_ims}
\end{figure*}

 The Clean algorithm with natural weighting is used to generate images using detectors 2 - 8, in energy bands 3-6, 6-12 and 12-25~keV. (Note, the flare occurred three months after RHESSI's second anneal. Prior to this anneal, use of detector 2 had not been recommended because it was not segmenting properly, but following the anneal it recovered this capability and was included in our imaging analysis.) Integration times were 60s, starting at 18:05:30, 18:06:30, 18:07:30 and 18:08:30. In Figure ~\ref{fig:RHESSI_ims}, RHESSI contours are shown overlaid on AIA 304~\AA\ and AIA 131~\AA\ images taken at the centre of the integration period. The contours show the same absolute count values in the first three image, and increase by a factor 2 in images 4 and 5. This makes clear the development of the absolute and relative intensities in each wavelength band. As time progresses, the ribbons become brighter relative to the southern loops at all energies, and in the two lower energy channels the emission is relatively uniform along the ribbons. At 12-25~keV the emission has a concentration around [-490", 80"], not seen at lower energies. At early times the 12-25k~keV data are noisy, but a tendency for the sources to align with the ribbons is visible. There is a good alignment between the ribbons in the RHESSI images and the ribbons seen in AIA~304~\AA\ images. The corresponding 131~\AA\ images show a fuzzy structure as well as compact footpoints at the locations of the RHESSI contours, suggesting the ends of hot loops may also contribute to the RHESSI emission.

\section{Flare ribbon energy losses and energy input}
\subsection{Radiative losses}\label{sect:radloss}

\begin{figure}
\centering
\includegraphics[width=8cm]{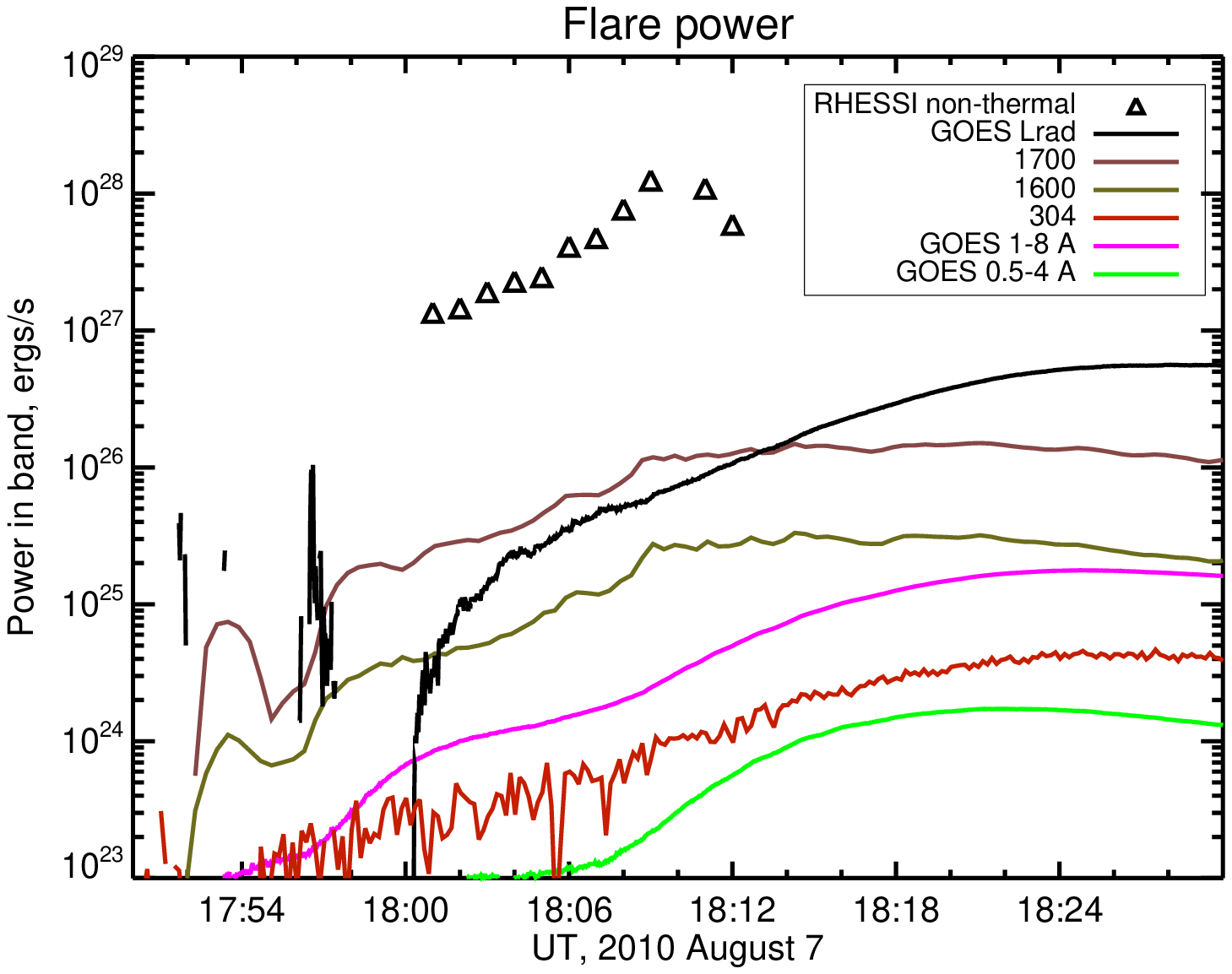}
  \caption{Power input and radiated from entire flare. Symbols show the RHESSI energy input rates, assuming a collisional thick target electron beam. Magenta and green are the GOES 1-8\AA\ and 0.5-4~\AA\ luminosities, and the black line is the calculated optically-thin radiation from the plasma detected by GOES. The red line is the excess EVE 304~\AA\ power, and the pink-brown and sage lines are the excess SDO/AIA 1700~\AA\ and 1600A~\AA\ powers.}
     \label{fig:luminosities}
\end{figure}

Figure~\ref{fig:luminosities} is a composite of the power radiated in several energy loss channels. From GOES we have the X-ray intensity ($\rm{L_{x}}$) in 1-8~\AA\ and 0.5-4~\AA\ channels, and the total optically-thin radiative losses ($\rm{L_{rad}}$) of the $\sim$ 10~MK plasma detected by GOES. This is provided by the SolarSoft GOES software, and calculation is based on the Chianti (Version 5) spectroscopic database \citep{2006ApJS..162..261L}. Coronal abundances are assumed. The flare excess power in the He II 304~\AA\ line measured by EVE is also shown, as is the power in the SDO/AIA 1600~\AA\ and 1700~\AA\ channels, calculated using the known telescope response functions in those channels.

In the first few minutes of the flare, until around 18:12~UT, the 1700~\AA\ passband dominates the flare radiative output. The 1700~\AA\ passband is designed to measure continuum, and has a passband of around 200~\AA. Full spectrum data are not available, and the event was not detectable in AIA/HMI data, presumably being too weak. So a total radiative power budget is hard to assemble, but it is reasonable to assume that the full UV-to-optical continuum will contain perhaps ten times the excess power in the 1700~\AA\ passband.
\cite{2011A&A...530A..84K} carried out a superposed epoch analysis of flares of different classes, finding that for flares in class C4 to M2.8 the the total irradiance is 250-500 times the GOES 1-8~\AA\ luminosity. This would correspond to $3.75 - 7.5 \times 10^{26}$ ergs~s$^{-1}$ at 18:06 UT and $5-10 \times 10^{26}$ ergs~s$^{-1}$ at 18:08 UT. The multiplying factor of 250-500 represents average values over the duration of the flare, however the longer wavelength emissions (i.e. optical to UV) which constitute the bulk of the radiated power, occur preferentially during the earlier parts of a flare.  An estimate of the ratio of the irradiance enhancement to the GOES 1-8~\AA\ luminosity enhancement during a flare's impulsive phase (as indicated by the maximum of the GOES derivative) was made in a flare reported by \cite{2004GeoRL..3110802W}. Measurements of  the X17 flare SOL2003-10-28:T11:10 show an increase in the total solar irradiance at the peak of the GOES derivative in SOL2003-10-28:T11:10 of 268ppm, or 360 mW/m$^2$, while the GOES flux at this time is around 1 mW/m$^2$. This gives a ratio of total power to GOES 1-8~\AA\  power during the impulsive phase of around 360, in the middle of the range found by \cite{2011A&A...530A..84K}. 

\subsection{Energy input from RHESSI}\label{sect:input}
The RHESSI data provide the means to estimate the energy input to the chromosphere, under the collisional thick target assumption that the hard X-rays are produced by collisional bremsstrahlung during the passage of non-thermal electrons through a dense target, in which the electrons are stopped completely by Coulomb collisions. The power delivered by non-thermal electrons above cutoff energy $E_c$ (in keV)  is estimated from the electron distribution parameters shown in Figure \ref{fig:rhessi_spec}, using:
\begin{equation}
P (E > E_c) = {{{\delta-1}\over{\delta-2}} {{10^{35}F_{35}}\over{E_c}} }\rm{erg~s^{-1}}
\end{equation}

where $F_{35}$ is the total number of electrons per second above the low-energy cutoff $E_c$, in units of $10^{35}$ electrons per second and $\delta$ is the electron spectral index.
The total power in electrons is shown by the triangle symbols in Figure~\ref{fig:luminosities}. There is sufficient energy to account for the radiated components at EUV and SXR, but as mentioned previously the energetically dominant UV-optical range is not sampled. From the arguments in Section~\ref{sect:radloss} we might expect the total power radiated to be $\sim 360$ times the GOES 1-8~\AA\ luminosity, or $\sim 10$ times the 1700~\AA\ power. This could readily be provided by a beam with the parameters inferred. There will also be conductive losses and work done to drive evaporation. We will return to the former in Section~\ref{sect:balance}.

\begin{figure}
\centering
\includegraphics[width=0.45\textwidth]{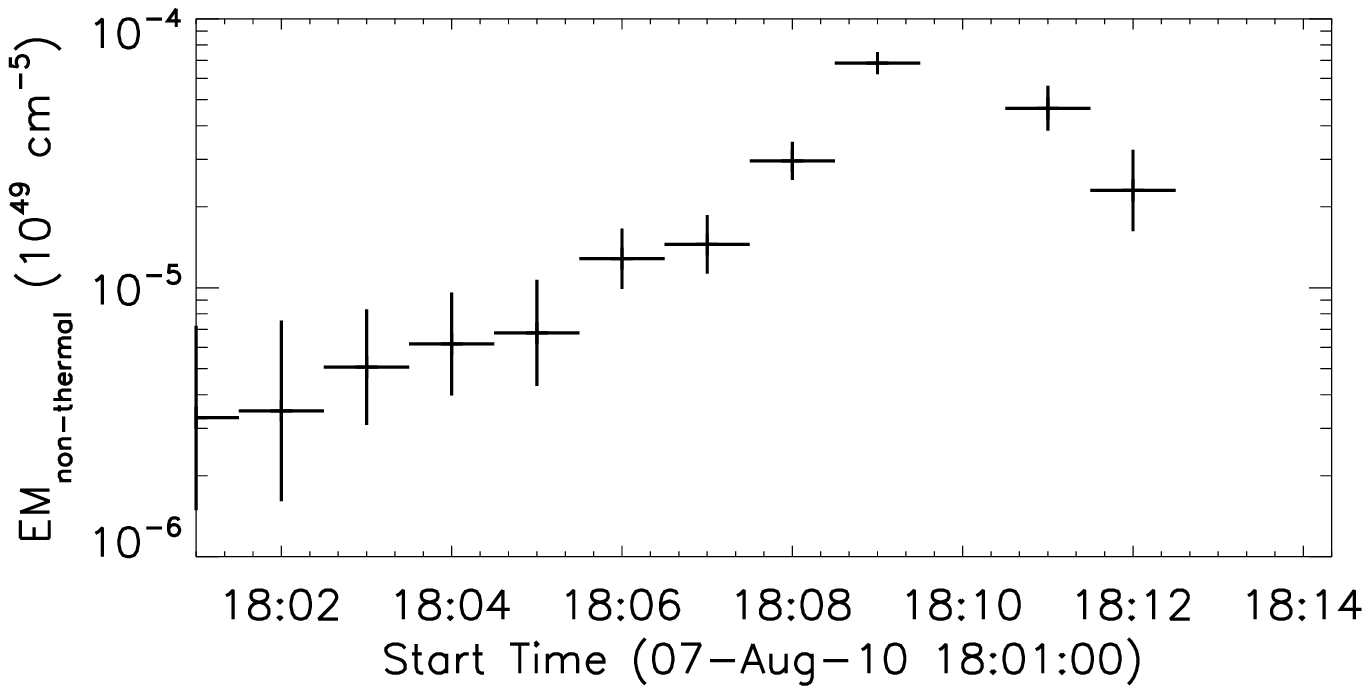}
\caption{Non-thermal EM for the event, calculated from the flare-integrated RHESSI spectrum. Error bars are calculated from errors on the fit to the photon spectral index.}
 \label{fig:emnt}
\end{figure}

The RHESSI data also allow us to evaluate the non-thermal EM, defined as:
$EM_{nt} = n_{nt} n_i V$,
where $n_{nt}$ is the number density of non-thermal electrons in the source
producing the HXR emission, $n_i$ the ambient ion density in that source, and $V$ the source volume. Knowing the thermal EM, $EM = n_e n_i V$ it is thus possible to determine the ratio of non-thermal to thermal electrons in the source. $EM_{nt}$ is determined from the fit parameters of the photon energy ($\epsilon$) spectrum $I{\epsilon} = A \epsilon^{-\gamma}$, e.g. \cite{1974SSRv...16..189L}:

\begin{equation}
EM_{nt} = 3.85 \times 10^{41} A \gamma (\gamma-1)^2 B\left(\gamma-{1\over 2}, {3 \over 2 }\right)\int_{E_{c}}^\infty E^{-\gamma + 1/2} dE,
\end{equation}

giving $EM_{nt}$ corresponding to non-thermal electron energy $E \ge E_{c}$. Here, $\gamma$ is the photon spectral index ($=\delta-1$ in the thick target chromosphere), and $B$ is the beta function. This gives the instantaneous number of non-thermal electrons in the radiating plasma. The relationship assumes that the HXR emission is due to collisional bremsstrahlung, but does not require the assumption of a thick-target. Again using the fit parameters shown in Figure~\ref{fig:rhessi_spec}, we arrive at the values for the non-thermal EMs shown in Figure~\ref{fig:emnt}. These values are $10^3$ to $10^4$ times lower than the typical thermal EMs $n_e n_i V$ obtained from RHESSI (Figure~\ref{fig:rhessi_spec}). In other words, averaging over all the plasma observed by RHESSI, one electron in $10^3 - 10^4$ has energy higher than the low-energy cutoff obtained from spectral fitting. The fraction of non-thermal electrons increases with time to a peak at some time between 18:09 and 18:11~UT.

\subsection{Power budget of the $\sim 10$ MK ribbon plasma}\label{sect:balance}
Here we examine in further detail the energy gains and losses from the 10~MK plasma in the ribbons identified with GOES, SDO/AIA and RHESSI. The total collisional power input $P_{\rm coll}~\rm{ergs~s^{-1}}$ to the heated part of the atmosphere only can be inferred from the collisional thick-target model by integrating the energy loss $\Delta E$ per particle transiting  over the electron distribution:
\begin{equation}
P_{\rm coll} (N) = \int_{E_c}^\infty F_o E_o^{-\delta}\Delta E (E_o, N) dE_o \label{eq:EN}
\end{equation}
where $N = \int n_e dl$ is the column depth of the heated plasma, and $\Delta E = E_o - E(N)$, with $E(N)$ being the energy of an electron of initial energy $E_o$ after it has traversed $N$. Subscript $o$ refers to the properties at injection into the chromosphere, i.e. the collisional thick target properties determined from spectral fitting.  From \cite{1978ApJ...224..241E}, $E(N)^2 = E_o^2 - 3\Lambda KN$, where $\Lambda$ is the Coulomb logarithm and $K = 2 \pi e^4$, $e$ being the charge of the electron. This approximation assumes that electrons enter as a unidirectional beam along the magnetic field direction (assumed vertical) and we take $\Lambda = 20$ which is appropriate for the fully ionized plasma of the upper chromosphere. 

Substituting for $\Delta E$ in Equation (\ref{eq:EN}) and performing the integral gives
\begin{equation}
P_{\rm coll}(N) = F_{\rm tot}E_c{{\delta-1}\over{\delta-2}}\left[ 1 - {\left({{\delta\over 2} - 1}\right) x_c^{1 - \delta/2} B\left(x_c; {\delta\over 2} -1, {3\over 2}\right)} \right]
\end{equation}\label{eq:eloss}
where
\begin{equation}
x_c = {{3KN}\over{E_c^2}} \label{eq:xc}
\end{equation}
and $B$ is now the incomplete Beta function. We  also use
\begin{equation}
F_{\rm tot} = \int_{E_c}^\infty F_o E_o^{-\delta}dE_o
\end{equation}
where $F_{\rm tot}$ is the total number of electrons per second injected above $E = E_c$. 

$N$ can be estimated from the footpoint (column) EM as $N \sim n_e L \sim (EM~L)^{1/2}$, for a uniform source of thickness $L$. The AIA ribbon sources in all channels co-align to better than one or two pixels, so that if the $10$~MK plasma sits above the AIA sources at 304~\AA\ then the maximum thickness of the $10$~MK plasma should be $\sim$ 1000~km (comparable also to the thickness of the chromosphere). Using the observed EM of $\sim 5\times 10^{28}{\rm cm}^{-5}$ gives a density $n_e \sim 2\times10^{10}{\rm cm}^{-3}$, with column depth $N$ of $\sim 2\times 10^{18}{\rm cm^{-2}}$. Assuming a smaller source thickness of $\sim$ 100~km results in $n_e \sim 7\times10^{10}{\rm cm}^{-3}$ and column depth $\sim 7\times 10^{17}{\rm cm^{-2}}$. 

\begin{figure}
\centering
\includegraphics[width=0.45\textwidth]{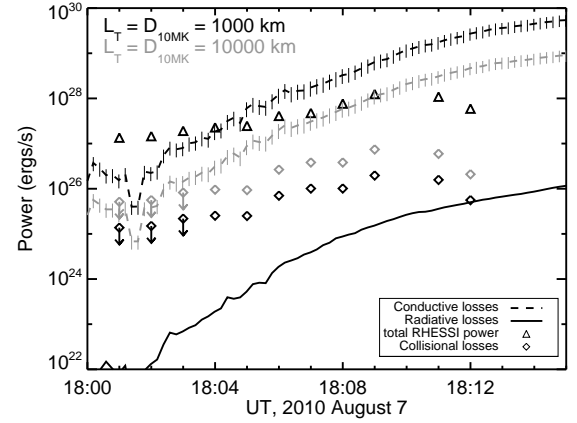}
\caption{Total non-thermal power delivered to the flare (triangles) and the collisional power input to the 10~MK plasma only (diamonds) compared to the radiative and collisional losses from the 10~MK plasma, assuming values of 1,000~km (black) and 10,000~km (grey) for the thickness of the 10~MK plasma/temperature scale-length. Error bars on $L_{\rm{cond}}$ are shown for an uncertainty of 1~MK in the temperature of the hot plasma }
 \label{fig:ebalance}
\end{figure}

The calculated values of $P_{\rm coll}$ for a source thickness $L$ = 1000~km are shown by the diamonds in Figure~\ref{fig:ebalance}. In this figure we have also applied a scaling to account for the fact that not all of the HXR emission comes from the ribbons. We have estimated the fraction of the non-thermal electron power delivered to the northern ribbons by multiplying the total power by the ratio of 12-25~keV flux summed over the northern box (shown in Figure~\ref{fig:131contours}) to the total 12-25~keV flux in the RHESSI image. 

Decreasing the thickness of the 10~MK source will decrease the collisional power input to the source. For completeness we also show losses for the case where the source thickness is increased to 10,000~km.

The hot  plasma emits optically thin radiation at rate $L_{\rm{rad}}~{\rm erg~s^{-1}}$, as discussed in Section~\ref{sect:radloss}, and we obtain the ribbon contribution by again scaling the total $L_{\rm{rad}}$ by the fractional 131~\AA\ intensity from the northern ribbon sources  (c.f. Fig~\ref{fig:AIA131curves}). The hot plasma must also lose energy by conduction to cooler, lower layers of the atmosphere at rate $L_{\rm{cond}}~{\rm erg~s^{-1}}$. The conductive losses are estimated following the approach of \cite{2009A&A...498..891B} for flux-limited conduction. Flux-limited conduction applies because the chromospheric temperature scale-length $L_T$ is small, referring again to the small separation between the 10~MK plasma imaged by AIA~131~\AA\ and 304~\AA\ channels. So the electron collisional mean-free path $l_{\rm{mfp}} = 5.21\times 10^3 T^2/n_e$ is significant compared to the temperature scale length. Non-local effects become important and classical Spitzer conductivity will not apply. A modification to the normal heat-flux equation is necessary, with the conductive flux $F_{\rm{cond}}$ (erg~cm$^{-2}$~s$^{-1}$) given by

\begin{equation}
F_{\rm{cond}} = \varrho(x) F_{\rm class} = \varrho(x)\kappa_o T^{5/2}{dT\over{dz}},
\end{equation}

where the classical Spitzer conductivity constant $\kappa_o$ is reduced by a factor $\varrho$, which is a function of $x = \log(l_{\rm{mfp}}/L_T)$, the logarithm of the ratio of the electron mean-free path to the temperature scale length. \cite{2009A&A...498..891B} fitted a functional form to the values of $\varrho$ published by \cite{1984PhRvA..30..365C} obtaining:

\begin{equation}
\varrho(x) = 1.01 e^{-0.05(x+6.63)^2}.
\end{equation}

Approximating $dT/dz$ by $T/L_T$, and taking $T = 10$~MK and $L_T = 10^3$~km we can calculate the conductive flux, using also the value of $n_e$ determined from the EM. With these values, the modification due to non-local conduction reduces $F_{\rm cond}$ by a factor 10 over the interval 18:00~UT to 18:12~UT, compared to the classical value.  $L_{\rm{cond}}$ is then obtained by multiplying $F_{\rm{cond}}$ by the total area of the hot northern ribbon sources observed in the 131~\AA\ channel. The results for radiative and conductive losses, and energy input, are shown in Figure~\ref{fig:ebalance}.

These calculations demonstrate firstly that conductive losses dominate radiative losses from the 10~MK ribbon plasma, and secondly that collisional energy losses in the 10~MK ribbon plasma, inferred from the collisional thick target calculation, are inadequate to supply the calculated conductive losses, by a factor of about 100 in the case of a source thickness of 1,000~km. However, the low-energy cut-off value $E_c$ on which the calculation depends is notoriously difficult to fit. If the value of $E_c$ is reduced to a fraction $f$ of the fitted value then the arrival rate of non-thermal electrons is increased by $f^{(1-\delta)}$, and the power delivered by $f^{(2-\delta)}$. A value of $f=0.4 - 0.5$ (i.e. $E_c$ around 4-6~keV) would thus provide adequate input power to balance the losses (the total beam power would also be increased). However $E_c$ is then close to the typical thermal energy of electrons in the $10$~MK plasma and it is not clear that `thermal' and `non-thermal' populations can be separated. Such a low cut-off energy also means that the cold collisional thick target approximation is not applicable to the whole distribution (see Section~\ref{sect:disc}.)

We note also that if the calculation is repeated for the case of electrons injected at some non-zero pitch-angle to the magnetic field, the effect is to introduce a factor $\mu$ (equal to the pitch-angle cosine) on the denominator of Equation (\ref{eq:xc}). The corresponding increase in $x_c$ results in an increase in $P_{\rm coll}$. For example, for $\mu = 0.3$, values of  $P_{\rm coll}$ typically increase by a factor 2.

\section{Discussion}\label{sect:disc}

This event provides a good opportunity to study the earliest phases of a large flare, in particular investigating the thermal and non-thermal properties of the flare ribbons. By demonstrating that the GOES and SDO/AIA 131\AA\ lightcurves track each other closely, and using AIA image data, we identify an interval between $\sim$ 18:05:00~UT and 18:10:00~UT during which a significant or even dominant fraction of the GOES emission comes from the flare ribbons, originating in or just above the chromosphere. We infer from GOES and AIA imaging that the ribbons contain $\sim 10$~MK plasma with a column EM of typically a few times $10^{28}{\rm cm}^{-5}$. This is backed up by AIA DEM analysis carried out on the flare ribbons, and also by RHESSI imaging and spectroscopy, which shows significant or dominant contributions to the thermal emission from the ribbons at this time. The conclusion is that the lower atmosphere is very strongly heated during the early phase of the flare. This is consistent with previous (but mostly overlooked)  observations of so-called impulsive soft X-ray footpoints \citep[e.g.][]{1993ApJ...416L..91M,1994ApJ...422L..25H,2004A&A...415..377M}.

In common with previous studies of the pre-flare we find that the initial brightening, in the southern loops, is located in a different place from the main flare ribbons. The southern loops dominate in RHESSI until $\sim$ 18:06~UT and cause a distinct bump in the GOES and SDO/AIA 131~\AA\ intensity. After 18:05~UT the ribbon emission starts to become important, and over the next few minutes there follows a strong increase in the inferred electron flux, and decrease in the electron spectral index. The typical column EM in the ribbons increases by about a factor two between 18:00~UT and 18:10~UT, which could correspond to the heating of deeper, denser layers of the chromosphere as the energy deposition rate increases. The increase in the total EM from the ribbons is caused mostly by an increase in the volume of hot plasma as the ribbons area increases.

The HXR spectrum from the entire flare region can be fitted by a thermal source with temperature approximately 10~MK, and a steep non-thermal spectrum with a low energy cutoff varying between 12 and 13~keV. Using standard thick-target expressions to interpret the non-thermal HXR spectrum and - by assuming the relationship between 1-8~\AA\ GOES X-ray luminosity and the flare total solar irradiance found by \cite{2011A&A...530A..84K} - it appears that the energy present in the non-thermal electrons can account for the totality of the radiated emission to be expected from this event. 

In addition, the energy budget of the 10~MK plasma of the northern ribbons is studied, comparing the collisional energy input to this plasma with radiative and conductive energy loss rates. Here it is found that to support the calculated conductive losses, the low-energy cutoff of the electron spectrum should be around 4 to 6~keV. Enthalpy flux from this plasma, i.e. chromospheric evaporation into overlying loops, is not considered but would place additional demands on the energetics.

A low-energy cutoff of only a few times the typical thermal energy of $kT \sim$ 1keV has two consequences. Firstly, the injected electrons near the cutoff would be in the warm-target regime $kT \lesssim E \lesssim 5kT$ meaning that they  experience both energy loss and energy gain in their interactions with the surroundings. Secondly the existence of two separate distributions (i.e. thermal and non-thermal) becomes questionable. 
\cite{2003ApJ...595L.119E} assesses the flare energetics under the collisional thick-target assumptions, including the effect of the warm-target, in the case where instead of a low-energy cutoff the power-law part of the spectrum continues smoothly into the thermal component. In such a scenario, and for electron spectral indices in the range $\delta = 6 - 9$, as we find in this flare, the total injected electron power would be $5\times 10^4$ to $5\times 10^7$ times the collisional thick-target power determined for electrons above $E_c$ (shown in Fig~\ref{fig:luminosities}) or $\sim 10^{34} \rm{ergs}~{s}^{-1}$, which is quite unreasonable.

We may consider an alternative scenario, which does not involve the injection of non-thermal electrons from the corona, but supposes their acceleration {\emph{in situ}} in chromospheric plasma that has been heated by some other means. Together, the thermal and the non-thermal EMs determined for the event indicate that  approximately one electron in $10^4$ at an energy greater than the fitted low-energy cutoff of $\sim 10$~keV is required, if we assume that the thermal and non-thermal EMs are generated in the same plasma. For our estimated $n_e$, this would correspond to a non-thermal electron number density of $2\times 10^6~{\rm cm}^{-3}$. At the peak of the HXRs the requirement is one electron in $10^3$. In principle, energy could be deposited in the plasma by some means other than electron collisions, for example by waves or turbulence, resulting first in electron heating and then in the acceleration of a high energy tail \citep[e.g.][]{2009ApJ...701L..34L}. We do not specify here what that heating/acceleration mechanism might be, but possibilities do exist - for example electron Landau resonance in Alfv\'enic turbulence, as explored by \cite{2010A&A...519A.114B}. The electron spectrum will then be determined by both the acceleration rate as a function of electron energy and the collisional loss rate. It is worth pointing out that if electron energy losses are dominated by collisions, the instantaneous {\emph{power}} in the non-thermal electron spectrum required to explain the non-thermal bremsstrahlung emission is the same as in the standard collisional thick-target model, however the {\emph{number}} of non-thermal electrons required to produce the emission can be significantly reduced, basically because any local acceleration increases the effective radiating lifetime of an individual electron compared to the case where its energy evolution is determined by collisional losses alone. More discussion of this can be found in \cite{2009A&A...508..993B}. 

The pre-flare chromosphere is so cool, dense and collisional that any energy input will initially result  only in heating. The electron-electron collision timescale for electrons in the thermal core of a Maxwellian distribution of temperature $T$ and density $n_e$ is \citep[e.g. ][]{2006pafp.book.....S}

\begin{equation}
\tau_{ee} = {{ m_e^2 (3 k T_e /m_e)^{3/2} }\over{8\pi e^4 n_e \Lambda} }{1\over0.714},
\end{equation}
where $m_e$ is the electron mass and $e$ the elementary charge. So as the plasma heats the collision timescale increases as $T^{3/2}$. For electrons of energy $E$ in the tail of the Maxwellian, the electron-electron collision timescale increases to 
$\tau_{\rm tail} = \chi^{3/2}\tau_{ee}$, where $\chi = E/kT$. So eventually some electrons out in the tail will become collisionless, in the sense that for these electrons the energy-input timescale becomes shorter than $\tau_{\rm tail}$. These electrons may then be accelerated. 

The fraction of electrons in the tail of a Maxwellian, above some energy $E = \chi  kT$, is

\begin{equation}
f_\chi = {\left(4\pi\over \chi\right)^{1/2}} e^{-\chi}.
\end{equation}

for $\chi >>$ 1. Observations require an accelerated fraction $f_\chi = 10^{-3}$, meaning that $\chi = 7.1$. If $f_\chi = 10^{-4}$ then $\chi = 9.3$. For a Maxwellian core temperature of 10~MK and a density of $2 \times 10^{10}\rm{cm}^{-3}$ we find $\tau_{ee} \sim 0.02$~seconds, therefore to accelerate the electrons in the tail would require an energisation process operating in the chromosphere with energy input timescale to these electrons of  $\lesssim \tau_{\rm tail} = 0.02 \times \chi^{3/2} \sim 0.02 \times 9^{3/2} \sim 0.5$ seconds.

We remark that any collisional losses undergone by the locally accelerated electrons cannot contribute substantially to the heating. The instantaneous energy density in the $10$~MK plasma is $n_e k T \sim 28~{\rm{erg~cm}^{-3}}$ for $n_e \sim 2 \times 10^{10}\rm{cm}^{-3}$, whereas the instantaneous energy density in the non-thermal electrons is $\sim 10^{-3}n_e E$ where $E$, the energy of the electrons, is typically 10~keV, giving an energy density in the non-thermals of $\sim 0.3~{\rm{erg~cm}^{-3}}$.

A variant on this scenario could involve the injection of a small number of non-thermal electrons from the corona, which are locally re-accelerated in chromospheric turbulence or current sheets \citep{2009A&A...508..993B}. But the requirement to identify a source of heating for the 10~MK ribbon plasma, other than collisional losses, is the same. As mentioned before, both the EUV emission and the RHESSI X-rays in the main ribbons appear to be weaker in regions of weaker magnetic field. We have not examined this in any detail but the HMI information on magnetic field, coupled with the ability to examine the AIA DEM on a pixel-by-pixel basis \citep{2012arXiv1212.5529H} will enable us to investigate possible correlations in detail.

Flares like this in SDO/AIA with a long, well-observed onset are relatively rare, since usually image saturation quickly becomes a problem. However, numerous events are expected to be unsaturated at least for the first few seconds of the impulsive phase, and it will be interesting to see whether 10~MK ribbon plasma is confirmed in other SDO events.

\begin{acknowledgements}
We are grateful to an anonymous referee for insightful comments that led to improvements in our analysis and presentation. LF and HSH would like to thank the Max Planck Institute for Solar System Research for travel support and hospitality. SDO data come courtesy of NASA/SDO and the AIA, EVE, and HMI science teams, and we are grateful to NASA for its open data policy which enables work such as this. This research was supported by STFC rolling grant ST/I001808/1, by Leverhulme grant F00-179A and by EC-funded FP7 project HESPE (FP7-2010-SPACE-1-263086).
\end{acknowledgements}

\bibliographystyle{apj}
\bibliography{bib_07aug2010}

\end{document}